\documentclass[fleqn,usenatbib]{mnras}

\usepackage{newtxtext,newtxmath}

\usepackage[T1]{fontenc}

\DeclareRobustCommand{\VAN}[3]{#2}
\let\VANthebibliography\thebibliography
\def\thebibliography{\DeclareRobustCommand{\VAN}[3]{##3}\VANthebibliography}

\usepackage{graphicx}	
\usepackage{amsmath}	
\usepackage{threeparttable}
\usepackage{bm}
\usepackage{array}   
\usepackage{booktabs} 
\usepackage{multirow} 
\usepackage{caption}
\usepackage[skip=0.5ex]{subcaption}
\usepackage{ulem} 

\newcommand{\angstrom}{\textup{\AA}}

\title[Accretion Disk RM in Close Binary SMBHs]{Continuum Reverberation Mapping of Accretion Disks Surrounding Supermassive Black Hole Binaries: Observational Signatures}

\author[Y.-X. Fu et al.]
{Yi-Xin Fu,$^{1,2}$ Yan-Rong Li,$^{1}$\thanks{E-mail: liyanrong@mail.ihep.ac.cn (YRL)}
Jian-Min Wang,$^{1,3,4}$\thanks{E-mail: wangjm@mail.ihep.ac.cn (JMW)}
Keith Horne,$^5$ 
Juan V. Hernández Santisteban,$^5$
\newauthor 
Roberta Vieliute,$^5$
Rick Edelson,$^6$
Tingting Liu,$^7$
Michael S. Brotherton,$^8$
Luka \v{C}. Popovi\'c,$^{9,10}$
\newauthor
Andjelka B. Kova\v{c}evi\'c,$^9$
and 
Shuo Zhai$^4$
\\
$^{1}$State Key Laboratory of Particle Astrophysics, Institute of High Energy Physics, Chinese Academy of Sciences, 19B Yuquan Road, Beijing 100049, China\\
$^{2}$University of Chinese Academy of Sciences, 19A Yuquan Road, Beijing 100049, China\\
$^{3}$School of Astronomy and Space Sciences, University of Chinese Academy of Sciences, Beijing 100049, China\\
$^{4}$National Astronomical Observatories of China, Chinese Academy of Sciences, 20A Datun Road, Beijing 100020, China\\
$^{5}$SUPA Physics and Astronomy, University of St Andrews, Scotland, KY16 9SS, UK\\
$^{6}$Eureka Scientific Inc., 2452 Delmer Street Suite 100, Oakland, CA 94602, USA\\
$^{7}$Department of Physics and Astronomy, Georgia State University, 25 Park Place, Suite 605, Atlanta, GA 30303, USA\\
$^{8}$Department of Physics and Astronomy, University of Wyoming, Laramie, WY 82071, USA\\
$^{9}$University of Belgrade-Faculty of Mathematics, Department of Astronomy, Studentski trg 16, Belgrade, Serbia\\
$^{10}$Astronomical Observatory, Volgina 7, 11000 Belgrade, Serbia
}

\date{Accepted XXX. Received YYY; in original form ZZZ}

\pubyear{2025}

\begin{document}
\label{firstpage}
\pagerange{\pageref{firstpage}--\pageref{lastpage}}
\maketitle


\begin{abstract}
It has remained challenging to reliably identify sub-parsec supermassive black hole binaries (SMBHBs), despite them being expected to be ubiquitous. We propose a new method using multi-band continuum reverberation mapping to identify low-mass-ratio SMBHBs in active galactic nuclei. The basic principle is that, due to the presence of a low-density cavity between the mini-disks and the circumbinary disk, the continuum emissions show a deficit at certain wavelengths, leading to a distinguishing feature in the relation between the inter-band time lag and wavelengths $\tau(\lambda)$. Specifically, the relation appears flat at short wavelengths because of the truncated sizes of the mini-disks and transits to a power law $\lambda^{4/3}$ at long wavelength stemming from the circumbinary disk. This transition feature is distinct from the uniform relation $\lambda^{4/3}$ of the standard accretion disk around a single black hole. Using the lamp-post scenario and assuming that only the secondary black hole is active in a low-mass-ratio SMBHB, we design a simple continuum reverberation model to calculate the transfer function of the accretion disks and the resulting $\tau(\lambda)$ relations for various SMBHB orbital parameters. The transition  wavelength typically can lie at UV/optical bands, mainly depending on the total mass and orbital separation of the SMBHB. We apply our SMBHB model to the intensive multiwavelength monitoring data of the SMBHB candidate PG1302-102 and find that the SMBHB model can reproduce the inter-band time lags. Remarkably, the inferred total mass and orbital period from the SMBHB fitting are consistent with values derived from other independent methods.
\end{abstract}

\begin{keywords}
accretion, accretion disks -- black hole physics -- quasars: supermassive black holes
\end{keywords}



\section{Introduction}
In the current hierarchical paradigm of galaxy formation and evolution, galaxies undergo frequent mergers with a redistribution of dark matter halos and baryonic matter (e.g., \citealt{Cole2000}). During this process, the supermassive black hole (SMBH) residing at the center of each progenitor galaxy is driven into the nuclear region of the newly formed galaxy due to dynamic friction exerted by the surrounding medium (e.g., \citealt{Colpi2014}). A gravitationally bound system forms once the separation between the two SMBHs come close enough so that their orbital velocity exceeds the stellar velocity dispersion (e.g., \citealt{Begelman1980}). Those close SMBH binaries (hereafter SMBHBs) with a separation less than 0.1~pc have drawn great interest because they shed light onto the so-called ``final-parsec problem'' (e.g., \citealt{Merritt2005}) and also constitute important emitting sources for nano-Hertz gravitational waves (e.g., \citealt{Sesana2004, Chen2020}).

Although they are expected to be prevalent across the universe, identifying those SMBHBs through electromagnetic radiations has always been challenging and elusive because of their small spheres of gravitational influence, which means that at large distances, their dynamical influence on the surrounding environment is indeed indistinguishable from those of single SMBHs. Over the past decades, various electromagnetic methods have been proposed to search for SMBHBs (see \citealt{DOrazio2023} for a recent review) and a number of possible candidates are claimed through systematic searches (e.g., \citealt{Graham2015, Charisi2016, Liu2019, Chen2020, Chen2024, Luo2025}) or serendipitous discoveries by different studies (e.g., \citealt{Valtonen2008, Bon2012, Yan2015, Li2016, Li2019, Zhen2016, ONeill2022, Abdollahi2024}).

Numerical simulations show that the SMBHB's tidal torques make certain regions of the accretion flow unstable and can open up a cavity in the flow (\citealt{Artymowicz1994, Farris2014, Shi&Krolik2015,Bowen2019,Lopez2021,Combi&Lopez2022}). The cavity is deprived of gas and thereby effectively gives rise to an emission deficiency around the wavelength corresponding to the typical temperature of the accretion flow in the same region. Such a cavity manifests as an observable decrement or notch in the spectral energy distribution (\citealt{Gultekin2012,Roedig2014}). However, the spectral energy distributions can be easily attenuated by the internal gas and dust distributed in the sources, leading to ambiguity in the interpretation (e.g., \citealt{Yan2015, Leighly2016}). 

In this work, we demonstrate a new observational signature of the cavity in the accretion disks surrounding SMBHBs. Previous intensive monitoring of active galactic nuclei shows that the accretion disk surrounding single SMBHs displays tight correlations between UV and optical continuum variability (\citealt{Starkey2017, Cackett2018, Cackett2020, Edelson2019, Kara2021, Kara2023, Edelson2024, Prince2025}). The inter-band time lag increases with wavelength, typically following the $\tau\propto\lambda^{4/3}$ power-law relation as anticipated from the standard accretion disk model and the lamp-post scenario (\citealt{Cackett2022}). As for the accretion disks surrounding SMBHBs, if the tight inter-band correlations continue to hold, the appearance of a cavity leads to deficit responses at certain wavelengths, resulting in an imprint on the inter-band continuum reverberations. The transfer functions of the accretion disks and the relation between the inter-band time lag and wavelength are expected to differ in a distinctive way from those of single SMBHs. This provides an alternative way to identify SMBHB candidates from modern time-domain databases.
Using a simple SMBHB model with the lamppost scenario, we calculate in detail the continuum reverberation mapping~(RM) and study how the reverberation signals depend on the SMBHB's orbital parameters.

This paper is organized as follows. In Section \ref{sec_method}, we describe the methodology for calculating inter-band reverberations of accretion disks in an SMBHB system. In Section \ref{sec_results}, we summarize the results. In Section~\ref{sec_pg1302}, we apply the SMBHB model to the intensive multiwavelength monitoring data of the quasar PG~1302-102, an SMBHB candidate initially identified based on the evidence for periodic variations by \cite{Graham2015}. We present discussions in Section \ref{sec_discussions}, followed by conclusions in Section~\ref{sec_conclusions}.

\section{Method}\label{sec_method}
\begin{figure*}
    \includegraphics[width=0.8\textwidth]{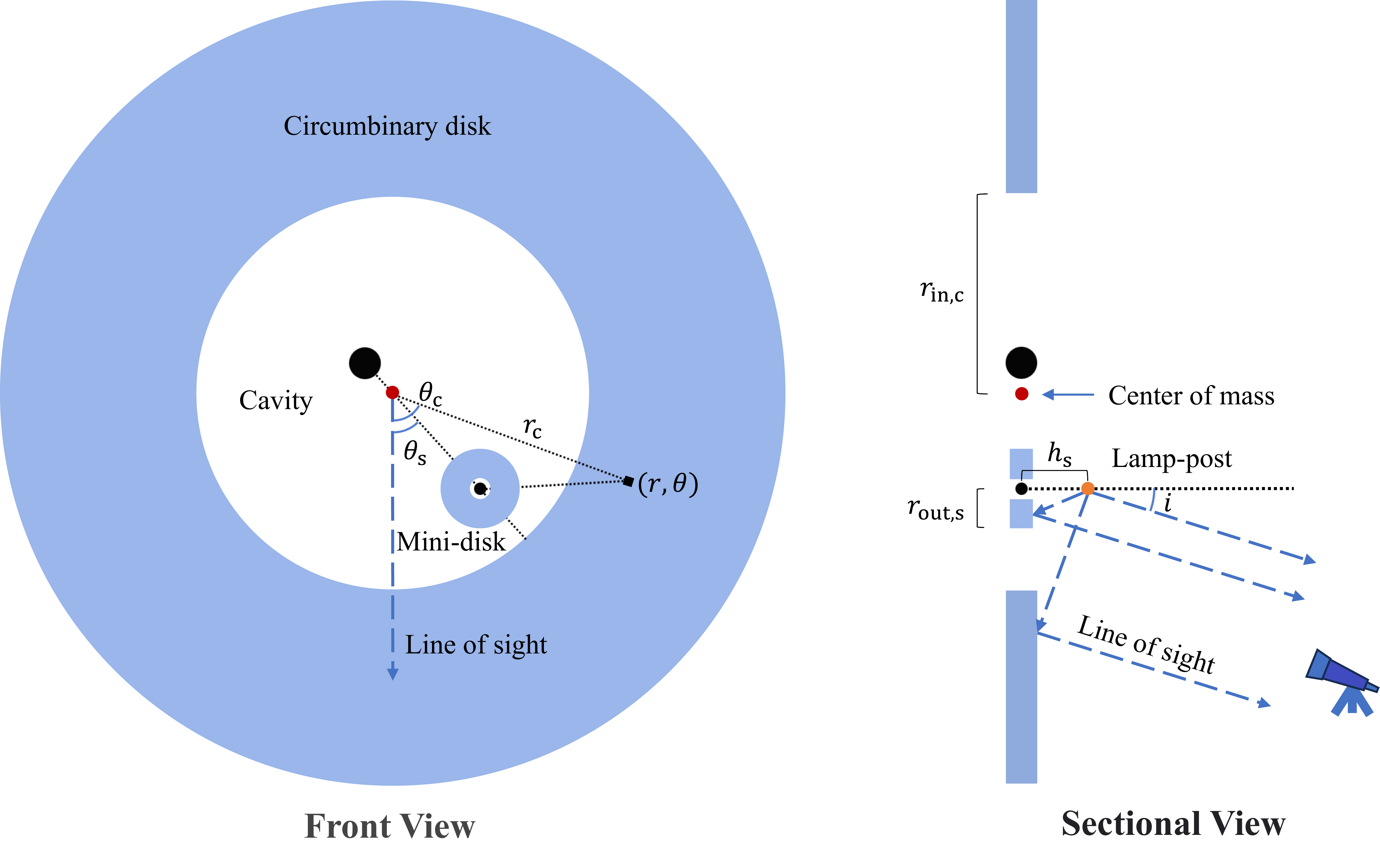}
    \caption{A schematic for accretion disks around an SMBHB (not to scale). The left panel shows the face-on view and the right panel shows the cutaway view. The binary black holes move around each other with a circular orbit. The secondary black hole carries a mini-disk, which has an outer radius $r_{\rm out, s}$. The circumbinary disk is coplanar with the mini-disk and has an inner radius $r_{\rm in, s}$. Between the mini-disk and circumbinary disk exists a low-density cavity. The orange point represent the location of the lamp-post source, which is above the mini-disk with a height of $h_{\rm s}$. The SMBHB is viewed at an inclination of $i$.}
    \label{fig:SMBHBs}
\end{figure*}

\subsection{Accretion disk model surrounding SMBHBs}
We start with the following simple configuration of accretion disks around SMBHBs (see Fig.~\ref{fig:SMBHBs}). Each black hole retains a mini-disk around itself, which is truncated by the tidal torque from the companion black hole and therefore has an outer edge limited by the respective Roche radius. Around the SMBHB, there exists a circumbinary disk, which feeds gas to the mini-disks through tidal streams. The region between the mini-disks and circumbinary disk has a low gas density and forms a cavity (\citealt{Artymowicz1996, Dorazio2013,Farris2014}).  

For the cases of unequal mass ratio, previous numerical simulations showed that accretion occurs preferentially onto the secondary black hole because it orbits closer to the inner rim of the circumbinary disk and therefore is more easily able to acquire gas from the circumbinary disk (\citealt{Farris2014,Shi&Krolik2015,Bowen2019,Combi&Lopez2022}). For the sake of simplicity, we neglect the mini-disk around the primary black hole and consider only the emissions from the mini-disk around the secondary black hole as well as from the circumbinary disk. This simplification is reasonable for the mass ratio of the SMBHB from a few percent to 0.25 (e.g., \citealt{Roedig2012,Farris2014}).


We denote the mass of the primary and secondary black holes as $M_{\rm p}$ and $M_{\rm s}$, respectively. The total mass is $M_{\rm t}=M_{\rm p}+M_{\rm s}$ and the mass ratio is $q=M_{\rm s}/M_{\rm p}$. The binary has a circular orbit with a semi-major axis $a_{\rm B}$ and period $P_{\rm B}$. 

The mini-disk and circumbinary disk are assumed to be geometrically thin and coplanar with the binary's orbital plane. The two disks are both optically thick so that the radiation at each annulus follows the Planck function,
\begin{equation}\label{Blackbody}
 B(\lambda ,T)= \frac{2hc^2}{\lambda ^{5}}\frac{1}{e^{hc/ \lambda kT}-1},
\end{equation}
where $T$ is the surface temperature of the accretion disk, $h$ is the Planck constant, $c$ is the speed of light, and $k$ is the Boltzmann constant.
The inner boundary of the circumbinary disk can be approximated as (e.g., \citealt{Gultekin2012,Yan2015})
\begin{equation}\label{eqn_r_inc}
 r_{\rm in,c} \sim a_{\rm B}/(1+q)+r_{\rm H,s},
\end{equation}
where the first term in the right-hand side represents the distance from the secondary black hole to the center of mass and the second term 
\begin{equation}\label{eqn_r_hs}
r_{\rm H,s}=a_{\rm B}(q/3)^{1/3},
\end{equation}
is the Hill radius of the secondary black hole (\citealt{Gultekin2012}).  We set the outer boundary of the circumbinary disk large enough to ensure the corresponding characteristic wavelength of emitted photons much longer than the wavelength range under investigation.
The outer radius of the mini-disk around the secondary black hole is limited by the Roche-lobe size, which is calculated as \citep{Eggleton1983}
\begin{equation}\label{eqn_r_Rs}
r_{\rm R,s}= \frac{0.49 a_{\rm B} q^{2/3}}{0.6q^{2/3} + \ln(1 + q^{1/3})}.
\end{equation}
By taking into account the possibility that the accreting gas might not fill up the entire Roche lobe, we set the outer radius of the mini-disk as a fraction of the Roche-lobe size,
\begin{align}\label{eqn_r_outs}
 r_{\rm out, s}=\xi r_{\rm R, s},
\end{align}
where $\xi < 1$ (\citealt{Artymowicz1994}). The inner radius of the mini-disk is set to be  the innermost stable circular orbit of a Schwarzschild black hole, namely, $r_{\rm in, s}=6R_{\rm g,s}$, where $R_{\rm g,s}=GM_{\rm s}/c^2$. 

\subsection{Continuum reverberation model}\label{sec: RM model}
The lamp-post model is widely used to explain tight correlations between ultraviolet and optical variability observed in active galactic nuclei (e.g., \citealt{Cackett2021}). In this model, there is an X-ray variable source, commonly referred to as a corona, seated at several gravitational radius above the SMBH, perpendicular to the disk plane. The variable X-ray photons emitted from this corona source illuminate the accretion disk. A fraction of photons are reflected and the other fraction are absorbed and thermalized by the accretion disk, leading to an increase in the disk's surface temperature. This effectively causes the thermal ultraviolet/optical radiation from the disk to co-vary in time with the illuminating X-ray flux, but with a time lag due to the light travel time from the corona to the accretion disk.

In the context of SMBHBs, this lamp-post model
describes the variability of emissions from both the mini-disk and circumbinary disk. Specifically, the corona above the secondary black hole (see Fig.~\ref{fig:SMBHBs}) illuminates both the mini-disk and circumbinary disk and drives the observed variability in ultraviolet/optical wavelength. 
For the convenience of calculations, we create a cylindrical coordinate ($r,\theta,z$) with the origin attached to the secondary black hole. The corona has a coordinate of ($0, 0, h_s$). This coordinate frame indeed moves along with the orbital motion of the secondary black hole. We use $\theta_{\rm s}$ to denote the orbital phase angle of the secondary black hole (see Fig.~\ref{fig:SMBHBs}) and let the line $\theta=0$ to be always parallel with $\theta_{\rm s} =0$. Also, we set the line of sight to have an inclination $i$ and its projection onto the orbital plane parallel with the line $\theta_{\rm s}=0$.

For a surface element of the mini-disk at $\bm{r}=(r,\theta)$, it is easy to show that the time lag between the radiation from this element and that from the corona is given by
\begin{equation}\label{eqn_taus}
    \tau_{\rm s}(r, \theta, i)= \left(\sqrt{h_{\rm s}^{2}+r^{2}}+ h_{\rm s} \cos i -r\cos \theta\sin i\right)/c,
\end{equation}
where $i$ is the inclination of the observer's line of sight to the disk and $h_{\rm s}$ is the height of the corona. For the circumbinary disk, we simply treat it as axisymetric, with the origin at center of mass of the SMBHB. For the convenience of calculations, we create a polar coordinate ($r_c, \theta_c$) for the circumbinary disk with the origin at the center of mass. A surface element of the circumbinary disk at $\bm{r_{\rm c}}=(r_{\rm c}, \theta_{\rm c})$ has a time lag dependent on the location of the secondary black hole, which is given by
\begin{equation}\label{eqn_tauc}
\tau_{\rm s}(r_c, \theta_c, i, \theta_s)=
\left(\sqrt{h_s^2 + d_{\rm c}^2}+h_s\cos i - d_{\rm p}\sin i\right)/c,
\end{equation}
where 
\begin{equation}
d_{\rm c} = \sqrt{r_c^2+\frac{a_{\rm B}^2}{(1+q)^2}-\frac{2a_{\rm B} r_c}{1+q}\cos(\theta_s-\theta_{\rm c})},
\end{equation}
and 
\begin{equation}
d_{\rm p} = r_c\cos\theta_c - \frac{a_{\rm B}}{1+q}\cos\theta_s.
\end{equation}

By taking into account the illumination from the lamp-post and assuming that the corona emission is isotropic, the temperature distributions of the mini-disk and circumbinary disk are respectively written 
\begin{equation}\label{T}
T^{4}(t,r,\theta,i) = T_{\rm b}^4(t-\tau_{\rm s}) + T_{\rm vis}^4,
\end{equation}
where 
\begin{equation}\begin{split}
T_{\rm vis}^4 = \left \{
    \begin{aligned}
    &\frac{3GM_{\rm s}\dot{M_{\rm s}}}{8 \pi \sigma r^{3}}\left(1- \sqrt{\frac{r_{\rm in,s}}{r}}\right) &\quad {\rm for~mini-disk},\\
    &\frac{3GM_{\rm t}\dot{M_{\rm c}}}{8 \pi \sigma r_{\rm c}^{3}}
    &\quad {\rm for~circumbinary \ disk},\\
    \end{aligned}
\right.
\end{split}\end{equation}
and
\begin{equation}\label{Tb}\begin{split}
T_{\rm b}^4(t-\tau_{\rm s}) =&
\frac{L_{\rm b}(t- \tau_{\rm s})(1-a)h_{\rm s}}{4 \pi\sigma}\\
&\times\left\{
\begin{aligned}
&\left(r^{2}+h_{\rm s}^{2}\right)^{-3/2}  &\quad {\rm for~mini-disk,}\\
&\left(d_{\rm c}^{2}+h_{\rm s}^{2}\right)^{-3/2}  &\quad {\rm for~circumbinary \ disk.}\\
\end{aligned}
\right.
\end{split}
\end{equation}
where $\dot{M}_{\rm s}$ and $\dot M_{\rm c}$ are the accretion rate of the mini-disk and circumbinary disk, respectively. $L_{\rm b}(t)$ is the lamppost luminosity at time $t$, $a$ is the disk albedo. 
The factor $\left(1-\sqrt{r_{\rm in, s}/r}\right)$ for the mini-disk arises from the zero-torque assumption at the inner edge. This factor is negligible for the circumbinary disk. 

A combination of Equations (\ref{Blackbody}) and (\ref{T}) illustrate that the observed emissions of both the mini-disk and circumbinary disk at a given wavelength vary with time due to the variable corona luminosity. We can mathematically characterize the corona luminosity variability into a constant and variable component, namely,
\begin{equation}\label{Lb}
L_{\rm b}(t)= L_{\rm \mu}+ L_{\rm \sigma}  X_{\rm b}(t),
\end{equation}
where $L_{\rm \mu}$ and $L_{\rm \sigma}$ denote the long-term mean and standard deviation of $L_{\rm b}(t)$, and $X_{\rm b}(t)$ is a dimensionless stochastic time series with a zero mean ($\langle X_b \rangle =0$) and unity variance ($\langle X_b^2 \rangle =1$). Similarly, the disk flux at a wavelength $\lambda$ and time $t$ can be characterized into 
\begin{equation}\label{flux}
F(\lambda ,t) = {F}_{\rm \mu}(\lambda)+ F_{\rm \sigma}(\lambda) X(\lambda,t),
\end{equation}
where again $F_\mu(\lambda)$ and $F_\sigma(\lambda)$ denote the long-term mean and standard deviation. Here, as a result of response to the corona variations, the dimensionless stochastic time series $X(\lambda, t)$ can be regarded as a blurred and  delayed echo to $X_{\rm b}(t)$, namely, 
\begin{equation}\label{eqn_X}
X(\lambda,t) = \int _{\rm 0}^{\rm \infty}\psi( \lambda, \tau)X_{\rm b}(t- \tau)\mathrm{d}\tau,
\end{equation}
where $\psi(\lambda,\tau)$ is the transfer function that represents the contributions of the surface elements of both the mini-disk and cicumbinary disk to the observed flux variation at a wavelength $\lambda$ and time lag $\tau$, written as~\citep{Cackett2007, Starkey2017}
\begin{equation}\label{resp}
    \psi(\lambda, \tau) = A(\lambda) \iint \frac{\partial B(\lambda,T)}{\partial T}\frac{\partial T}{\partial L_{\rm b}}\frac{\partial L_{\rm b}}{\partial X_{\rm b}} \delta(\tau - \tau_{\rm s}) r\mathrm{d}\theta \mathrm{d}r,
\end{equation}
where $\delta(x)$ is the Dirac delta function and $A(\lambda)$ is a normalization factor to ensure\footnote{We note that since Equation~(\ref{eqn_X}) implies $\langle X^2\rangle < 1$, the quantity $F_\sigma(\lambda)$ in Equation~(\ref{flux}) over-estimates the standard deviation of the echo light curve $F(\lambda, t)$.
However, this effect is likely small in practice because AGN light curve variances are typically dominated by timescales longer than the width of the transfer function. Also, this effect does not affect the results of the present work, as the time-lag calculations are independent of $F_\sigma(\lambda)$.}
\begin{equation}
\int_{0}^{\infty}\psi(\lambda, \tau) \mathrm{d}\tau = 1.
\end{equation}
In Appendix~\ref{app:A}, we present a derivation for the expression of $\psi(\lambda, \tau)$.

The above defined $\psi(\lambda, \tau)$ is nomenclaturally called the responsivity-weighted transfer function in the literature as it reflects the responsivity to variations of the driving light curve (e.g., \citealt{LW2024}). Another type of transfer function is called the emissivity-weigthed transfer function, defined by the integral of the emissions (i.e., $B(\lambda, T)$) over the whole disk surface.
It is argued that in RMs, the responsivity-weighted transfer function is more relevant to the cross-correlation analysis mentioned below, which is sensitive to variability in light curves (e.g., \citealt{LW2024}).
Throughout the calculations, we use the responsivity-weighted transfer function in subsequent calculations. 
 
There are two factors meriting remarks. First, along with the orbital motion of the secondary black hole, the orbital phase angle $\theta_s$ changes accordingly, resulting in a time dependence of reverberation of the circumbinary disk. 
Provided the orbital period is much longer than the period length of light curves for reverberation analysis, it is still reasonable to assume that the coordinate frame is stationary in the calculations.
Second, due to the variability of $L_{\rm b}$, the surface temperatures of both mini-disk and circumbinary disk are a function of time and thereby their reverberation are also subject to fluctuations.  
To facilitate computations, we simply replace $L_{\rm b}$ with $L_\mu$ to calculate the surface temperature in Equation (\ref{T}). We expect that over a long period, the short-timescale fluctutations of the corona luminosity might be blurred out and effectively, we  would obtain averaged time lags that are consistent with those derived using the mean corona luminosity. This is supported by the linear UV/optical flux-flux relations observed in disk RM campaigns (e.g., \citealt{Starkey2016, HS2020, Donnan2023, Edelson2024, Prince2025}).

\begin{table*}
    \centering
    \caption{The fiducial parameter set. }
    \label{tab:para}
    \begin{tabular}{cccl} 
		\hline
		Parameter & Unit & ~~~~~~~~~~Value~~~~~~~~~~ & Note\\
		\hline
		$M_{\rm t}$ & $10^8 M_{\odot}$ & $5.25$ & Total mass of the SMBHB\\
		$M_{\rm p}$ & $10^8 M_{\odot}$ & $5$ & Mass of the primary SMBH\\
		$M_{\rm s}$ & $10^8 M_{\odot}$ & $0.25$ & Mass of the secondary SMBH\\
        $q$         & ... & 0.05 & Mass ratio ($M_{\rm s}/M_{\rm q}$) of the SMBHB \\
        $a_{{\rm B}}$ & $R_{\rm g,t}$ & $300$ & Separations of the SMBHB's orbit ($\sim 1500~{\rm AU}$)\\
        $P_{\rm B}$ & day & 977& Period of the SMBHB's orbit\\
		$\dot{M}_{\rm c}$ & $\dot{M}_{\rm Edd,c}$ & $0.1$ & Accretion rate of the circumbinary disk ($\sim 1.2~M_\odot~{\rm yr^{-1}}$) \\
        $\dot{M}_{\rm s}$ & $\dot{M}_{\rm Edd,s}$ & $2.1$ & Accretion rate of the mini-disk ($\sim 1.2~M_\odot~{\rm yr^{-1}}$)  \\
        $L_{\mu}$ & $\eta \dot{M}_{\rm s} c^2$ & $f_{\rm b}$ & Mean luminosity of the corona ($f_{\rm b } = 1.0$, $\eta=0.1$)\\
        $a$ & ... & 0.1 & Albedo of the disks  \\
        $h_{\rm s}$ & $R_{\rm g,s}$ & $6 $ & Height of the corona \\
        $r_{\rm s, in}$ & $R_{\rm g,s}$ & $6 $ & Inner radius of the mini-disk\\
        $r_{\rm s, out}$ & $r_{\rm R,s}$ & $0.1$ & Outer radius of the mini-disk ($\sim 100~R_{\rm g,s}$) \\
        $r_{\rm c, in}$ & $a_{\rm B}$ & $(1+q)^{-1} + (q/3)^{1/3}$ & Inner radius of the circumbinary disk ($\sim360~R_{\rm g,t}$) \\
        $r_{\rm c, out}$ & $R_{\rm g,t}$ & $10^5$ & Outer radius of the circumbinary disk \\
        $\theta_{\rm s}$ & degree & $0$ & Orbital phase angle of the SMBHB\\
        $i$ &  degree & $30$ & Inclination angle\\
        $T_{\rm GW}$\tnote{1} & yr & $3 \times 10^5$ & Gravitational wave radiation timescale \\
		\hline
    \end{tabular}
\end{table*}

\subsection{Calculating time lags}\label{sec:W-D}
From the obtained transfer function, we can directly determine the peak and centroid time lags. However, in realistic observations, we only obtain light curves at some wavelength bands, rather than the transfer function.
The time lag between two light curves can be measured via the cross-correlation function (CCF; \citealt{Gaskell1986, Gaskell1987}). In this method, two types of time lag are usually derived. One is the peak time lag corresponding to the maximum CCF ($r_{\rm max}$) and the other is the centroid time lag, conventionally defined as the centroid of the CCF above 80\% of $r_{\rm max}$ (\citealt{Gaskell1987}).
Mathematically speaking, there is a relation between the transfer function and CCF of two light curves, e.g., say, $X_{\rm b}(t)$ and $X(t)$  (\citealt{Maoz1991, Welsh1999})
\begin{equation}\label{eqn_ccf}\begin{split}
    {\rm CCF}(\tau, \lambda) &= \int X_{\rm b}(t) X(\lambda, t+\tau) \mathrm{d}t\\
    &= \int^\infty_0 \psi(\lambda, \tau'){\rm ACF}(\tau -\tau')\mathrm{d}\tau',
\end{split}\end{equation}
where ACF is the auto-correlation function (ACF) of $X_{\rm b}(t)$, defined as
\begin{equation}
{\rm ACF}(\tau)=\int X_{\rm b}(t) X_{\rm b}(t+\tau) \mathrm{d}t.
\end{equation}
As can be seen, the CCF of two light curves is deemed to be a convolution between the ACF of the driving light curve and the transfer function. As a result, the peak and centroid time lags from the transfer function and the CCF are not exactly equivalent. 
To surmount this issue, we employ the following procedure to calculate time lags from the transfer function.

Previous studies showed that AGN variability can be approximately described by the damped random walk (DRW) model (e.g., \citealt{Kelly2009, Li2013, Zu2013, Zhou2024}). 
A DRW process has a covariance function of 
\begin{equation}
    S(\Delta t)= \sigma_{\rm D} ^{2} \exp \left( -\frac{|\Delta t|}{\tau_{\rm D}}\right),
\end{equation}
where $\Delta t$ denotes the time difference of two points in the light curve and $\sigma_{\rm D}$ and $\tau_{\rm D}$ denote the variability amplitude and characteristic damping timescale, respectively. The ACF of a DRW process is thereby
\begin{equation}\label{ACF}
    {\rm ACF}(\tau ) = \exp \left( -\frac{|\tau|}{\tau_{\rm D}}\right).
\end{equation}

Given a calculated transfer function $\psi(\lambda, \tau)$, we first convolve it with the ACF by using Equations (\ref{eqn_ccf}) and (\ref{ACF}) and obtain a new function
\begin{equation}\label{eqn_tf_conv}
\psi_\otimes(\lambda, \tau) = \int_0^\infty \psi(\lambda, \tau')\exp \left( -\frac{|\tau-\tau'|}{\tau_{\rm D}}\right) \mathrm{d}\tau'.
\end{equation}
We adopt the value of $\tau_{\rm D}$ according to the empirical relation between $\tau_{\rm D}$ and black hole mass ($M_\bullet$) established by \cite{Lu2019}
\begin{equation}\label{eqn_taud}
    \log\left(\frac{\tau_{\rm D}}{{\rm day}}\right) = (1.56 \pm 0.37) + (0.60 \pm 0.09 )\log\left(\frac{M_\bullet}{10^8M_\odot}\right),
\end{equation}
which applies to $V$-band AGN variability.
From $\psi_\otimes(\lambda, \tau)$, we then derive the peak time lag $\tau_{\rm peak}$ and calculate the centroid of the entire time lag range  as 
\begin{equation}\label{tau_cent}
    \tau_{\rm cent} 
    = \int^\infty_0 \tau\psi_\otimes(\lambda, \tau)\mathrm{d}\tau \bigg/ \int^\infty_0 \psi_\otimes(\lambda, \tau)\mathrm{d}\tau.
\end{equation}
It is easy to prove that $\tau_{\rm cent}$ equals to the centroid of $\psi(\lambda, \tau)$.
To align with the CCF recipe mentioned above, we also calculate the centroid of the region above 80\% of the maximum value of $\psi_\otimes(\lambda, \tau)$, 
\begin{equation}\label{tau_cent2}
\widetilde\tau_{\rm cent} 
    = \int_{>80\%} \tau\psi_\otimes(\lambda, \tau)\mathrm{d}\tau \bigg/ \int_{>80\%} \psi_\otimes(\lambda, \tau)\mathrm{d}\tau.
\end{equation}

It is worth stressing that there exist other more sophisticated, but model-dependent methods to determine time lags and/or transfer functions of light curves (e.g., \citealt{Horne1994,Li2016MICA, Li2021, Anderson2021}). In particular, an inference of the transfer function from observed light curves allows us to directly compare with the theoretical transfer function from our SMBHB model.  However, such an application of these methods is quite beyond the scope of this paper and will be deferred to future works.

\section{Results}\label{sec_results}
In this section, we firstly showcase the transfer function and resulting relation between wavelength and time lag ($\tau(\lambda)$) for a fiducial set of SMBHB parameters, and then explore the orbital parameter space of the SMBHB and discuss their influences on the $\tau(\lambda)$ relation. We find that a distinctive signature of SMBHBs is that the $\tau(\lambda)$ relation no longer obeys a power law (e.g., \citealt{Edelson2019, Cackett2021, Guo2022}), instead, it is generally divided into two power laws with a break at some specific wavelength $\lambda_{\rm tran}$. We present in Section~\ref{sec:lag break} an approximate formula for the dependence of $\lambda_{\rm tran}$ on orbtial parameters of the SMBHB .

\subsection{The fiducial case}\label{sec:fiducial}
We adopt a fiducial set of parameters to showcase our calculations. The values of parameters are summarized in Table~\ref{tab:para}. The SMBHB has a total mass of $M_{\rm t}=5.25\times 10^8M_{\odot}$, a mass ratio $q=0.05$, and an orbital semi-major axis of $a_{{\rm B}}=300 R_{\rm g,t}$, where $R_{\rm g,t}=GM_{\rm t}/c^2$. This configuration yields a mass of the primary black hole $M_{\rm p}=5\times 10^8M_{\odot}$ and the secondary black hole $M_{\rm s}=0.25\times 10^8M_{\odot}$.
The mass accretion through the circumbinary disk feeds the binary SMBHs. Considering that accretion occurs predominately onto the secondary SMBH in low-mass-ratio SMBHB (e.g., \citealt{Farris2014, Shi&Krolik2015, Bowen2019, Combi&Lopez2022}), we neglect the accretion onto the primary SMBH and simply set $\dot{M}_{\rm c} = \dot{M}_{\rm s}$. We adopt a dimensionless accretion rate of the circumbinary disk $\dot{m}_{\rm c} \equiv \dot{M}_{\rm c}/\dot{M}_{\rm Edd,c} = 0.1$, which corresponds to the characteristic rate observed in normal AGNs (e.g., \citealt{Rakshit2020,Su2025}).
As a result, the dimensionless accretion of the mini-disk is $\dot{m}_{\rm s} \equiv \dot{M}_{\rm s}/\dot{M}_{\rm Edd,s} = \dot m_{\rm c} (1+q)/q = 2.1$. Here, $\dot{M}_{\rm Edd,c} \equiv L_{\rm Edd,c}/\eta c^2$, $\dot{M}_{\rm Edd,s} \equiv L_{\rm Edd,s}/\eta c^2$ are the Eddington accretion rates and $L_{\rm Edd, c}=6.60\times 10^{46}~{\rm erg~s^{-1}}$ and $L_{\rm Edd, s}=0.31 \times 10^{46}~{\rm erg~s^{-1}}$ are the Eddington luminosites for the SMBH with a mass of $M_{\rm t}$ and $M_{\rm s}$, respectively. The radiative efficiency is set to $\eta=0.1$ (e.g., \citealt{Marconi2004}). 

The outer radius of the mini-disk around the secondary black hole is set to $r_{\rm out,s}=0.1 r_{\rm R, s}$ (see Equation~\ref{eqn_r_outs}). As demonstrated in Section~\ref{sec_pg1302}, this choice aligns with the inference from fitting our SMBHB model to the observed data of the SMBHB candidate PG 1302-102. 
The luminosity of the lamp-post is a fraction of the mini-disk's luminosity, i.e., $L_{\rm b}=f_{\rm b}\times(\eta \dot M_{\rm s}c^2)$. Throughout the calculations, we fix the fraction $f_{\rm b}=1$, which means that the the lamp-post has a comparable luminosity with the mini-disk. The height of the lamp-post is fixed to $h_{\rm s}=6R_{\rm g, s}$, consistent with the observational contraints from X-ray microlensing (\citealt{Chen2012, Morgan2012}) and X-ray reflection observations (\citealt{Chainakun2016,Epitropakis2016}).
The disks' albedo is fixed to $a=0.1$. Note  that $f_{\rm b}$ and $(1-a)$ are degenerate in determining the disks' surface temperature (see Equation~\ref{T}). With the above setting, the ratio between corona and local viscous heating of the mini-disk is
\begin{equation}\begin{split}
\frac{T_{\rm b}^4}{T_{\rm vis}^4} &= \frac{2}{3}\eta f_{\rm b} (1-a)\frac{h_{\rm s}}{R_{\rm g, s}}\frac{r^3}{(r^2+h_{\rm s}^2)^{3/2}}\\
&\approx 0.36 f_{\rm b}\left(\frac{\eta}{0.1}\right)\left(\frac{1-a}{0.9}\right)\left(\frac{h_{\rm s}}{6R_{\rm g,s}}\right),
\end{split}
\end{equation}
where the factor $\left(1-\sqrt{r_{\rm in, s}/r}\right)$ is neglected.
This implies that the corona and viscous heating contribute comparable amounts of energy to the mini-disk, which is a common choice in disk reverberation mapping analysis (e.g., \citealt{Fausnaugh2016, Starkey2016, Starkey2017}). We stress that the transition feature in the $\tau(\lambda)$ relation described below is quite insensitive to the choice of fiducial values.

\begin{figure}
    \includegraphics[width=1.0\linewidth]{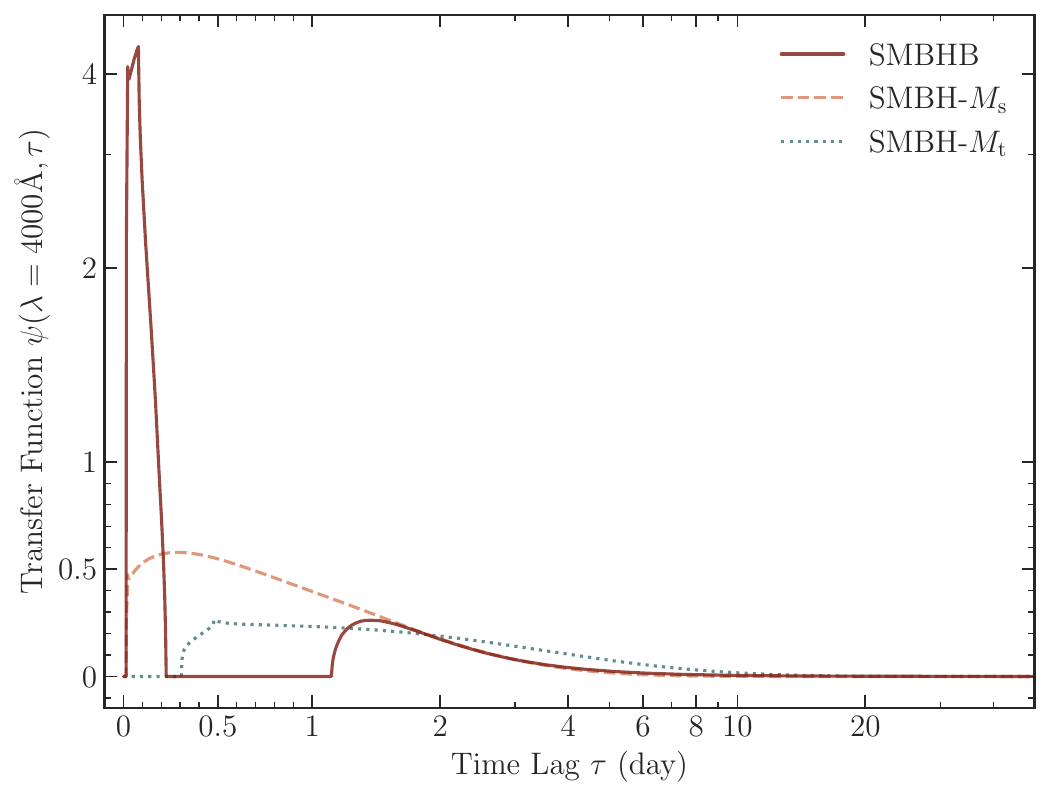}
    \caption{The responsivity-weighted transfer function of a SMBHB (red solid line) at 4000~{\AA} with the fiducial parameters listed in Table~\ref{tab:para}.
    The dotted and dashed lines indicate the transfer function for a single SMBH with a mass of $M_{\rm s}$ and $M_{\rm t}$, respectively. Note that the scales of both horizontal and vertical axes are adjusted to be ''symmetric logarithmic'', namely, linear between 0 and 1 and logarithmic beyond~1.}
    \label{fig:resp}
\end{figure}

\begin{figure*}
\centering
\includegraphics[width=1.0\textwidth]{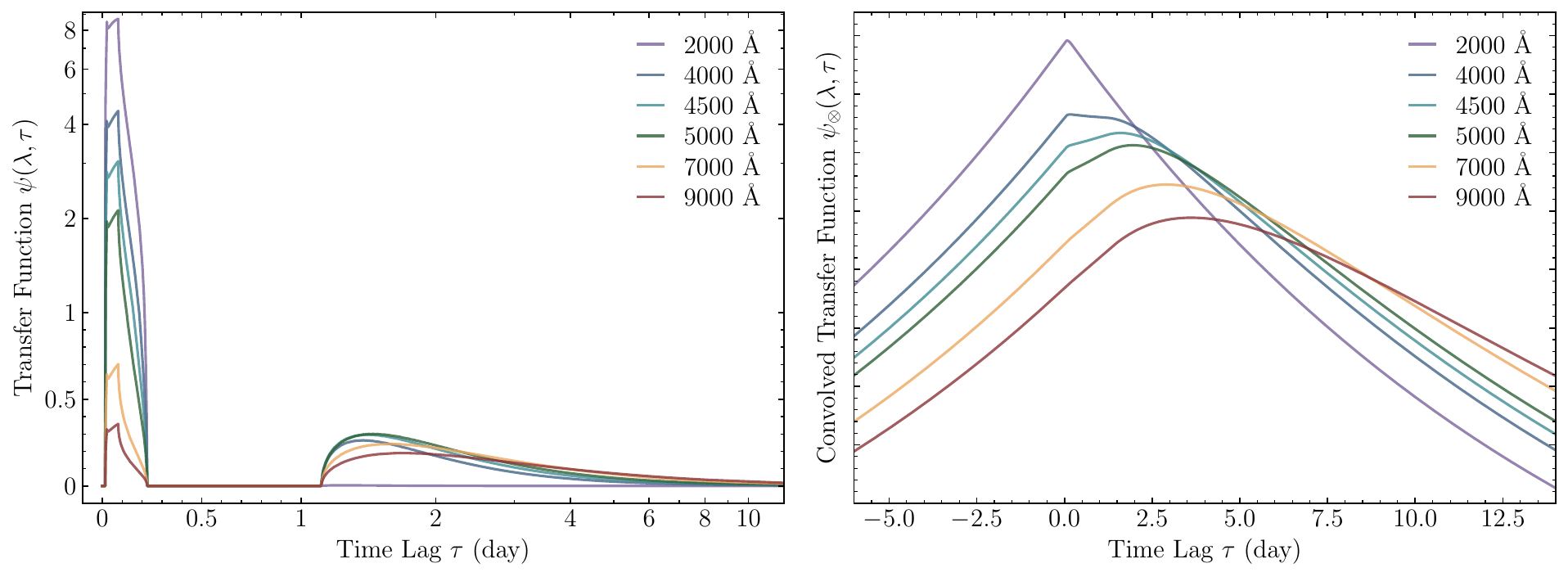}
\caption{(Left) The responsivity-weighted transfer functions of a SMBHB at different wavelengths. Note that the scales of both horizontal and vertical axes are adjusted to be ``symmetric logarithmic'', namely, linear between 0 and 1 and logarithmic beyond~1. (Right) The  transfer functions in the left panel convolved with an exponential ACF (see Equation~\ref{eqn_tf_conv}). The characteristic timescale of the ACF is set to $\tau_{\rm D}\approx 15$ days for $M_\bullet=0.25\times10^8M_{\odot}$ according to Equation~(\ref{eqn_taud}). The parameter values used for calculating the tranfer functions  are listed in Table~\ref{tab:para}.}
\label{fig:resp_wave}
\end{figure*}

\subsubsection{Responsivity-weighted transfer function}
In Figure.~\ref{fig:resp}, we show the calculated responsivity-weighted transfer function at $4000~\angstrom$ for our fiducial SMBHB. For the sake of comparison, we superimpose the transfer functions of the accretion disk around a single SMBH with a mass of $M_{\rm s}$ and $M_{\rm t}$, respectively. As can be seen, in contrast to those of a single SMBH, the transfer function of an SMBHB exhibits a bimodal structure, separated by a region of null response owing to the cavity.
The short-lag narrow response bump comes from the mini-disk, which is truncated at a time lag of $\sim r_{\rm s,out}(1+\sin i)/c\approx0.2$ days. The long-lag broad response distribution comes from the circumbinary disk, which starts at a time lag of $\sim r_{\rm H, s}(1-\sin i)/c\approx1.1$ days. As we will demonstrate below, such a bimodal structure makes the continuuum response of an SMBHB clearly distinguished from those of a single SMBH and therefore allows us to identify SMBHB candidates using multi-band time-domain survey data.

The left panel of Figure~\ref{fig:resp_wave} illustrates the responsivity-weighted transfer function at different wavelengths. As wavelength increases, the amplitude of the short-lag bump gradually declines whereas the amplitude of the long-lag response rises. At wavelength of 2000~{\AA}, the transfer function is dominated by the short-lag response from the mini-disk and the response from the circumbinary disk is negligible. This is because the mini-disk is much hotter than the circumbinary disk.

\subsubsection{The $\tau(\lambda)$ relation}\label{sec_relation}
As demonstrated in Section~\ref{sec:W-D}, to compare with measured time lags from CCF, one needs to convolve the transfer function with
the ACF of the driving light curve (see Equation~\ref{eqn_ccf}) to obtain $\psi_{\otimes}(\lambda, \tau)$ and then use it to calculate
peak and centroid time lags. The ACF can be approximated by an exponential function in Equation~(\ref{ACF}). We adopt the characteristic damping timescale $\tau_{\rm D}\approx 15$ days for $M_{\rm s}=0.25\times10^8M_{\odot}$ according to Equation~(\ref{eqn_taud}). 
In the right panel of Figure~\ref{fig:resp_wave}, we plot $\psi_{\otimes}(\lambda, \tau)$ at different wavelengths. The convolution causes $\psi_{\otimes}(\lambda, \tau)$ to no longer exhibit two distinct response peaks but rather to manifest as a single major peak with two minor servely blurred bumps. Importantly, as the wavelength varies from $4000~\angstrom$ to $5000~\angstrom$, the peak time lag has a sharp transition, shifting from the value associated with the short-lag bump to that of the long-lag bump.

\begin{figure*}
	\includegraphics[width=\textwidth]{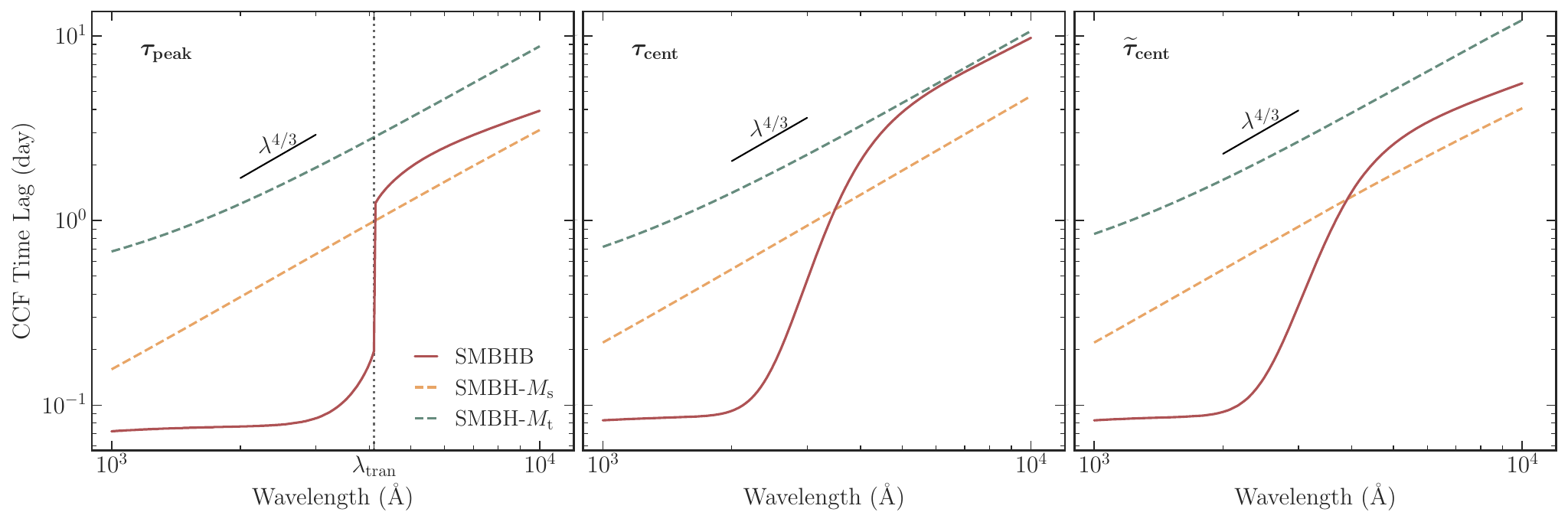}
    \caption{The $\tau(\lambda)$ relation of an SMBHB with the fiducial parameters listed in Table~\ref{tab:para}. From left to right panels are for time lags measured by the peak ($\tau_{\rm peak}$) and centroid ($\tau_{\rm cent}$) of the convolved transfer function, and the centroid ($\widetilde{\tau}_{\rm cent}$) of the convolved transfer function above 80\% of its maximum (see Equation~\ref{tau_cent2}), respectively. In each panel, the dashed lines represent the corresponding time lags for a single SMBH with a mass of $M_{\rm s}$ and $M_{\rm t}$, respectively. For $M_{\rm s}$ and $M_{\rm t}$, the transfer function is convolved with $\tau_{\rm D}=$ 15 and 100 days, respectively, according to Equation~(\ref{eqn_taud}). The power law with a slope of $4/3$ is plotted to guide the eye. In the left panel, the location of the lag break wavelength $\lambda_{\rm tran}$ is illustrated with grey dotted lines.}
    \label{fig:relation}
\end{figure*}

\begin{figure*}
\centering
    \includegraphics[width=1.0\textwidth]{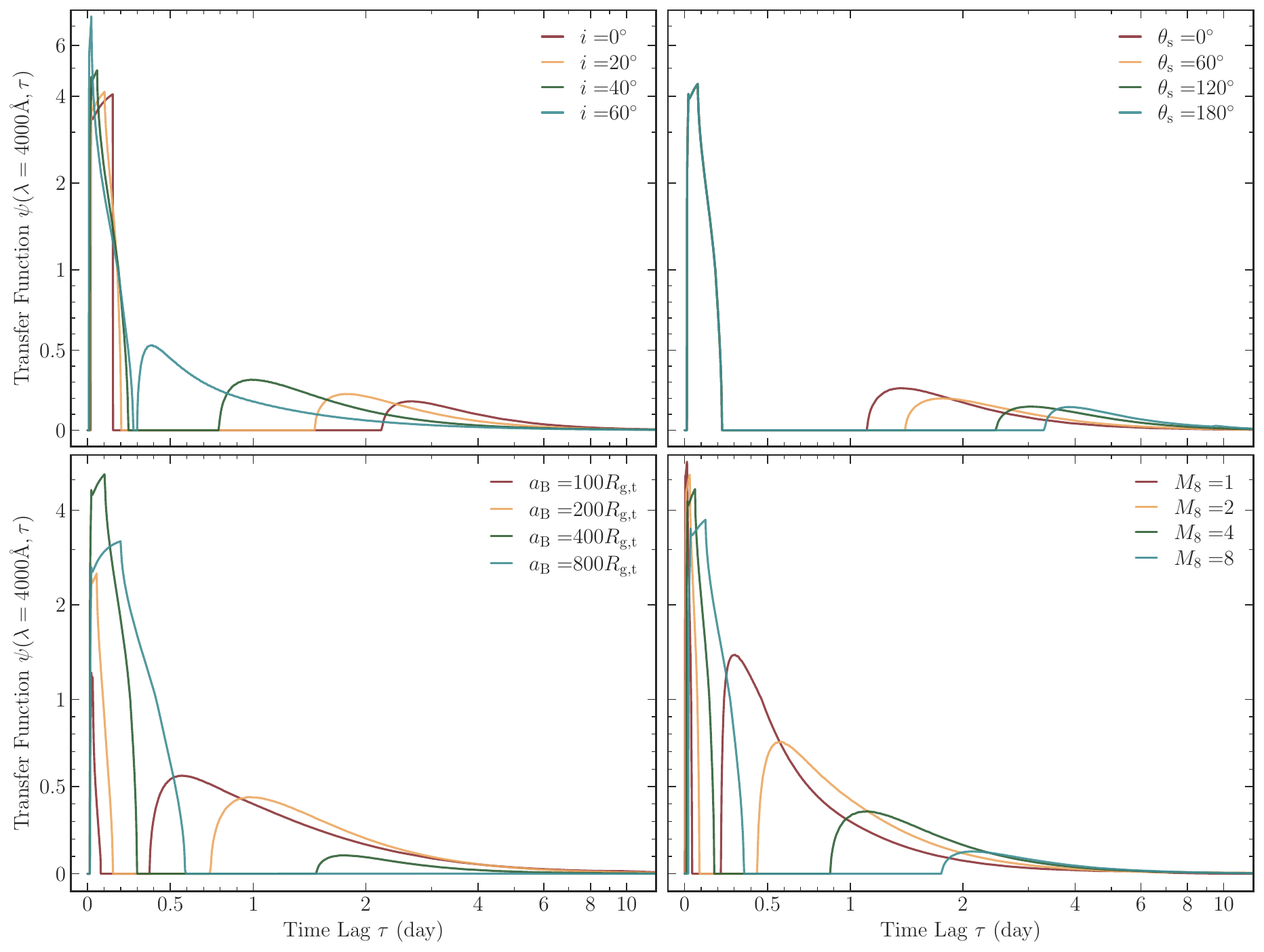}
    \caption{The responsivity-weighted transfer function for different inclinations (top left), phase angles (top right), orbital separation (bottom left), and total mass (bottom right). In each case, the rest parameters are set as the fiducial values listed in Table~\ref{tab:para}, and the wavelength is set to $4000~\angstrom$. Note that the scales of both horizontal and vertical axes are adjusted to be ``symmetric logarithmic'', namely, linear between 0 and 1 and logarithmic beyond~1. On the right-hand panel of the second row, we set $M_8=M_{\rm t}/10^8M_{\odot}$.}
    \label{fig:multi_resp}
\end{figure*}

Figure~\ref{fig:relation} shows wavelength dependence of the calculated peak and centriod time lags of the convolved transfer function as well as the centriod time lag of the convolved transfer function above 80\% of its maximum. For the sake of comparison, we superimpose the time lags for a single SMBH with a mass of $M_{\rm s}$ and $M_{\rm t}$. 
We find that the $\tau(\lambda)$ relations of SMBH conforms to a canonical power-law with a slope of $4/3$ (e.g., \citealt{Edelson2019, Cackett2021, Guo2022}).
For the SMBH with a mass of $M_{\rm t}$, the time lag at the short-wavelength end is slightly higher than the power-law expectation, due to the inner radius of the disk and the height of the lamp-post, resulting in a lower limit of several $R_{\rm g,t}/c$.

Due to the bimodal feature of the transfer function,
the $\tau(\lambda)$ relations of SMBHB for all three types of time lags deviate significantly from the $4/3$ power law. The relation displays a transition from a quite flat power law (index $\ll4/3$)  at shorter wavelengths to a relatively steeper power law (index $\sim4/3$) at longer wavelengths. The former corresponds to the response from the mini-disk, which is truncated at its outer edge (see Equation~\ref{eqn_r_outs}) due to the tidal torque of the primary black hole, resulting in time lags much shorter than those of a single SMBH. In addition, the limited outer edge also leads to the time lag being almost insensitive to wavelength.

The time lag at long wavelength mainly corresponds to the response from the circumbinary disk with a truncated inner edge (see Equation~\ref{eqn_r_inc}) and is shorter than that of a single SMBH with a same mass of $M_{\rm t}$. The reasons are twofold. Firstly, from Equation~(\ref{eqn_taus}), it follows that the earliest response of the circumbinary disk comes from the surface element at $r_c=r_{\rm in, c}$ and $\theta_{\rm c}=0^\circ$ (in our fiducial case of $\theta_{\rm s}=0^\circ$), approximately equal to $r_{\rm H, s}(1-\sin i)/c$ but smaller than $r_{\rm in, c}/c$, which roughly is the counterpart time lag for a single SMBH. This can also be clearly seen in Figure~\ref{fig:SMBHBs}. Secondly, the transfer function is bimodal and after a convolution in Equation~(\ref{eqn_tf_conv}), the effective time lag will be reduced due to the contribution from the short-time lag responses.

The transition in the $\tau(\lambda)$ relation  between short and long wavelength is the most evident when using $\tau_{\rm peak}$
(see the left panel of Figure~\ref{fig:relation}).
For $\tau_{\rm cent}$ or $\widetilde{\tau}_{\rm cent}$, because of the centroid implementation, the transition is not as dramatic as for $\tau_{\rm peak}$, but still clearly seen. As mentioned above, this transition is caused by the presence of the cavity with null responses. Accordingly, the wavelength of this transition is mainly controlled by the temperature of the circumbinary disk at the inner edge and the mini-disk at the outer edge. For the fiducial parameters listed in Table~\ref{tab:para}, the temperature at the outer edge of the mini-disk is $T\sim 5\times 10^4$ K, corresponding to a characteristic wavelength of $\lambda=2.9\times10^7\textrm{\AA}/T\sim 550$ {\AA}. The temperature at the inner edge of the circumbinary disk is $T\sim 6\times 10^3$~K, corresponding to $\lambda\sim$4900~{\AA}. In Figure~\ref{fig:relation}, we find that the transition wavelengths in all three types of time-lag measures lies around the latter wavelength. In Section~\ref{sec:lag break} below, we investigate how the transition wavelength depends on the orbitial parameters.

\begin{figure*}
    \includegraphics[width=0.88\textwidth]{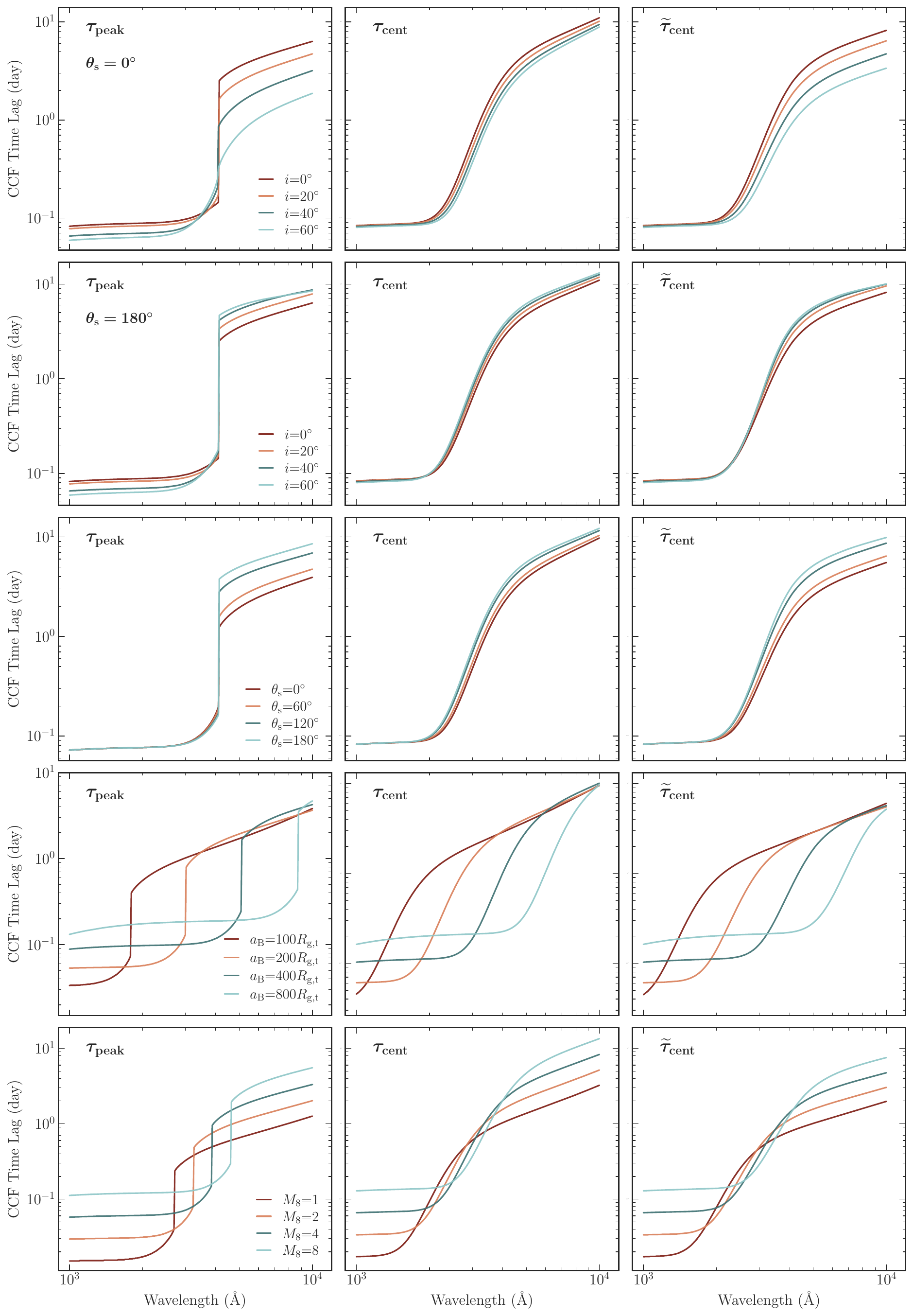}
    \caption{The $\tau(\lambda)$ relation of SMBHB for different inclinations (first and second row), phase angles (third row), orbit separation (fourth row), total mass (fifth row), mass ratio(sixth row) and outer radius of the mini-disk(seventh row) . In each case, the rest parameters are set as the fiducial values listed in Table~\ref{tab:para}, except for the second row, where we set $\theta_{\rm s}=180^\circ$.}
    \label{fig:lam-tau}
\end{figure*}
\begin{figure*}
  \ContinuedFloat
  \centering
  \includegraphics[width=0.88\textwidth]{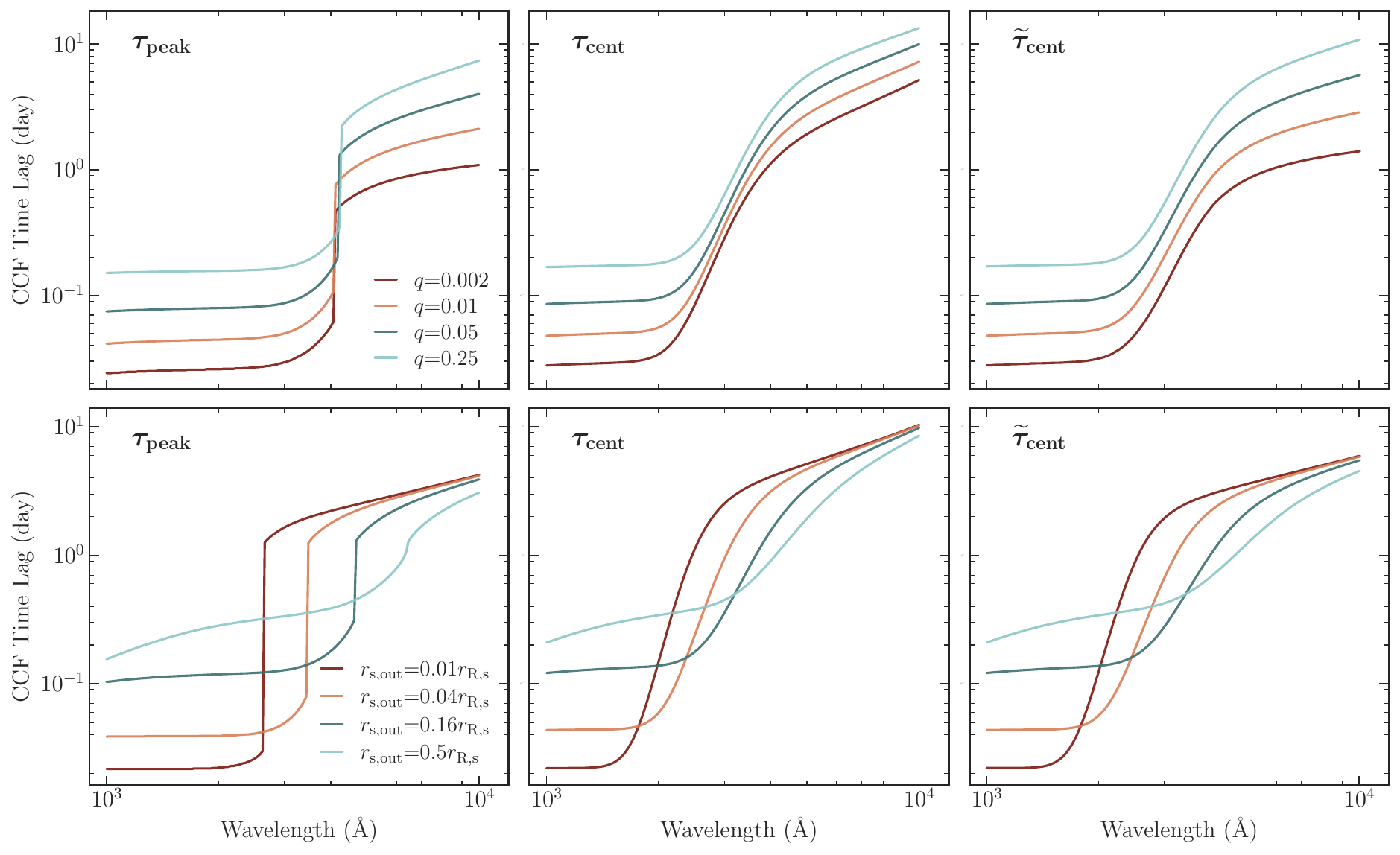}
  \caption{\it (Continued).}
  \label{fig:lam-tau-continued}
\end{figure*}

\subsection{Dependence of the responsivity-weighted transfer function on orbital parameters}\label{sec:transdependence}

All orbital parameters of an SMBHB system listed in Table~\ref{tab:para} affects to different degrees the shape of the responsivity-weighted transfer function.
In this section, we demonstrate the influences of the major parameters $i$, $\theta_{\rm s}$, $a_{\rm B}$, and $M_{\rm t}$. Figure~\ref{fig:multi_resp} shows the corresponding responsivity-weighted transfer function at $\lambda=4000~{\text{\AA}}$.

From Equations~(\ref{eqn_taus}) and (\ref{eqn_tauc}), it is straightforward to deduce that for the inclination increasing from 0$^\circ$ (face-on) to 60$^\circ$ (more edge-on), the lag of the truncated mini-disk becomes longer ($\propto(1+\sin i)$) whereas the starting lag of the circumbinary disk becomes shorter ($\propto(1-\sin i)$). As a result, the null-response region shrinks and the two response bumps from mini- and circumbinary disks gradually come together and overlap. Nevertheless, the shape of the transfer function is still distinct from that of a single SMBH (see Figure~\ref{fig:SMBHBs}).

The orbital phase angle $\theta_{\rm s}$ affects the response from the circiumbinary disk. As the secondary black hole moves from near side ($\theta_{\rm s}=0^\circ$) to far side ($\theta_{\rm s}=180^\circ$), the starting lag of the circumbinary disk changes from $r_{\rm H, s}(1-\sin i)/c$ to $r_{\rm H, s}(1+\sin i)/c$, increasing by a factor of three for $i=30^\circ$. Therefore, the separation between the two response bumps is enlarged.

The orbital separation $a_{\rm B}$ controls the outer radius $r_{\rm out, s}$ of the mini-disk in Equation~(\ref{eqn_r_outs}) and inner radius $r_{\rm in, c}$ of the circiumbinary disk in Equation~(\ref{eqn_r_inc}). The lag of the truncated mini-disk and the starting lag of the circumbinry disk increase with $a_{\rm B}$. The separation between the two response bumps scales with the Hill radius $r_{\rm H, s}$ in Equation~(\ref{eqn_r_hs}) and therefore also increases with $a_{\rm B}$. Meanwhile, as $a_{\rm B}$ increases, the circiumbinary disk recedes from the lamp-post in the secondary black hole and consequently, its response amplitude weakens.

The total mass $M_{\rm t}$ controls the physical scale of the SMBHB system. In our SMBHB configurations, the mass of the secondary SMBH $M_{\rm s}= q M_{\rm t}/(1+q)$ and the orbital separation $a_{\rm B}\propto R_{\rm g, t}\propto M_{\rm t}$. Hence, the size of both the mini-disk and circumbinary disk scales linearly with $M_{\rm t}$ and so does the time lag. Meanwhile, the disk surface temperature satisfies $T\propto M_{\rm t}^{-1/4}$ and the characteristic wavelength satisfies $\lambda\propto M_{\rm t}^{1/4}$ (see Equation~\ref{T}). As a result, if keeping other parameters unchanged, the shape of the transfer function $\psi(\lambda, \tau)$ is exactly identical under scaling of $\lambda$ by $M_{\rm t}^{1/4}$ and $\tau$ by $M_{\rm t}$. This in turn means that the $\tau(\lambda)$ relation maintains its shape while shifting along the direction of constant $\lambda^{-1}\tau^4$.

\begin{figure*}
\centering
    \includegraphics[width=\textwidth]{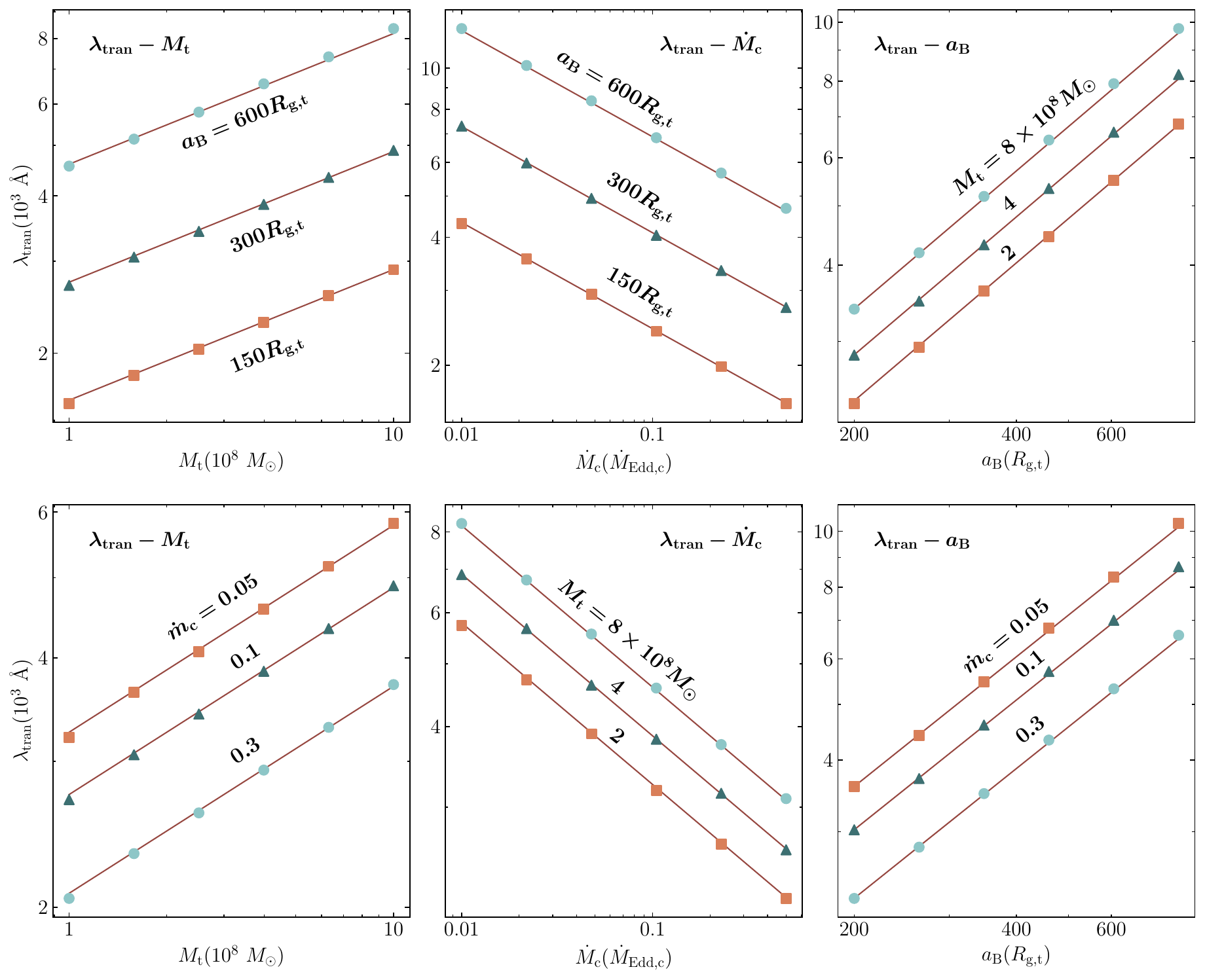}
    \caption{ The transition wavelength $\lambda_{\rm tran}$ from Equation~\eqref{eqn_lam_break} while varying (left) the total mass $M_{\rm t}$, (middle) accretion rate of the circumbinary disk $\dot{m}_{\rm c}$, and (right) the orbital separation $a_{\rm B}$. The text along with each line marks the value of the parameter under changing. The other parameters are set by the fiducial values listed in Table~\ref{tab:para}.}
    \label{fig:lambdac}
\end{figure*}

\subsection{Dependence of the $\tau(\lambda)$ relation on orbital parameters}\label{sec:dependence}

In Figure~~\ref{fig:lam-tau}, we illustrate the $\tau(\lambda)$ relation with different orbital parameters. The first and second rows are for the inclination in the cases of $\theta_{\rm s}=0^\circ$ and $\theta_{\rm s}=180^\circ$, respectively. The former corresponds to the case where the secondary SMBH is located on the near side whereas in the latter case the secondary SMBH is located on the far side.
In both cases, the transition wavelength $\lambda_{\rm tran}$ remains largely unchanged for different inclinations. However, it is interesting to note the opposite dependence of the time lag on the inclination at long wavelengths. For $\theta_{\rm s}=0^\circ$, the time lag decreases with the inclination, while for $\theta_{\rm s}=180^\circ$, the time lag increases with the inclination. This is because the minimum time lag of the circumbinary disk is $\tau\propto(1-\sin i)$ for $\theta_{\rm s}=0^\circ$ and $\tau\propto(1+\sin i)$ for $\theta_{\rm s}=180^\circ$.

The orbital phase angle $\theta_{\rm s}$ affects only the time lags of the circumbinary disk. The break wavelength is quite insensitive to $\theta_{\rm s}$. As $\theta_{\rm s}$ changes from $0^\circ$ to $180^\circ$, the secondary SMBH moves from the near side to the far side, resulting in the time lags of the circumbinary disk gradually increasing.

The fourth and fifth rows of Figure~~\ref{fig:lam-tau} illustrate the dependence of the $\tau(\lambda)$ relation on the orbital separation $a_{\rm B}$ and the total mass $M_{\rm t}$, respectively.
Both the outer radius of the mini-disk $r_{\rm out,s}$ and inner radius of the circumbinary disk $r_{\rm in,c}$ increase linearly with $a_{\rm B}$ (see Equation~\ref{eqn_r_outs} and Equation~\ref{eqn_r_inc}), resulting in a proportional increase in the time lags.
Meanwhile, the temperature at both the outer edge of the mini-disk and at the inner edge of circumbinary disk decrease with increasing $a_{\rm B}$. A combination of these two effects leads to a significant longward shift in the characteristic wavelength of $\lambda_{\rm tran}$ as $a_{\rm B}$ increases. The total mass affects the $\tau(\lambda)$ relation in a similar way as $a_{\rm B}$, since the total mass determines the physical scales of the SMBHB system. In particular, as mentioned in Section~\ref{sec:transdependence}, the $\tau(\lambda)$ relation exactly retains its shape under shifting along the direction of constant $\lambda^{-1}\tau^4$ when adjusting $M_{\rm t}$.

The sixth row of Figure~~\ref{fig:lam-tau} illustrates the impact of mass ratio $q$ on the $\tau(\lambda)$ relation. This mainly reflects on the following two aspects. The first aspect involves the modulation of the Roche and Hill radius by $q$, whereby the minimum distance from the inner edge of the circumbinary disk to the corona scales with $q^{1/3}$. Second, $q$ controls the mass of the secondary black hole, thereby determining the  the mini-disk's temperature and the corona height.
The joint effect of these two aspects leads to changes in time lag across all wavelengths. However, the transition wavelength $\lambda_{\rm tran}$ is insensitive to $q$ because the inner radius of the circumbinary disk $r_{\rm in, c}$ in Equation~(\ref{eqn_r_inc}) is dominated by the first term $a_{\rm B}/(1+q)$, which changes by only $\sim$20\% for $q$ varying from 0.002 to 0.25.

The last row of Figure~\ref{fig:lam-tau} illustrates the influence of the outer radius of the mini-disk $r_{\rm s, out}$. This parameter is introduced to account for the possibility that the Roche lobe of the secondary SMBH might not be fully filled. As $r_{\rm s, out}$ increases, the mini-disk extends in size, resulting in longer time lags at shorter wavelength. Meanwhile, the transition wavelength increases. When $r_{\rm s, out}$ reaches $0.5\,r_{\rm R, s}$, the lag-break feature in the $\tau(\lambda)$ relation is no longer prominent because the cavity between the mini-disk and the circumbinary disk shrinks and the null-response region in the transfer function diminishes.

\subsection{The transition wavelength $\lambda_{\rm tran}$}\label{sec:lag break}
In the preceding sections, we illustrate that the $\tau-\lambda$ relation of SMBHBs shows a transition in the slope from short to long wavelengths. The transition wavelength
approximates to the characteristic wavelength of the emissions at the inner edge of the circumbinary disk.
According to Equations~(\ref{T}) and (\ref{Tb}), the temperature at the inner edge of the circumbinary disk is not uniform, but changes with ($\theta_{\rm s}-\theta_{\rm c}$). Nevertheless, we can make an approximation as
\begin{equation}
\begin{split}
T_{\rm tran} &\approx \left(f\frac{3GM_{\rm t}\dot M_{\rm c}}{8\pi\sigma a_{\rm B}^3}\right)^{1/4}\\
&\approx 10^4 f^{1/4}\left(\frac{M_{\rm t}}{10^8M_\odot}\right)^{-1/4}\left(\frac{\dot m_{\rm c}}{0.1}\right)^{1/4}\left(\frac{a_{\rm B}}{100R_{\rm g,t}}\right)^{-3/4} {\rm K},
\end{split}
\end{equation}
where $f$ is a factor on the order of unity depending on the mass ratio $q$, the corona height $h_s$, and the corona luminosity ratio $f_{\rm b}$ as
\begin{equation}
f \approx \left[\frac{1}{1+q} + \left(\frac{q}{3}\right)^{1/3} \right]^{-3}\left[ 1 + \frac{2}{3}\frac{f_{\rm b}(1-a)h_s}{R_{\rm g, t}}\right].
\end{equation}
The characteristic wavelength is then given by
\begin{equation}\label{eqn_lam_break}
\lambda_{\rm tran} \approx 2700 f^{-1/4}\left(\frac{M_{\rm t}}{10^8M_\odot}\right)^{1/4}\left(\frac{\dot m_{\rm c}}{0.1}\right)^{-1/4}\left(\frac{a_{\rm B}}{100R_{\rm g,t}}\right)^{3/4} {\text{\AA}}.
\end{equation}
or
\begin{equation}\label{eqn_lam_break_p}
\lambda_{\rm tran} \approx 8600 f^{-1/4}\left(\frac{M_{\rm t}}{10^8M_\odot}\right)^{-1/4}\left(\frac{\dot m_{\rm c}}{0.1}\right)^{-1/4}\left(\frac{P_{\rm B}}{\rm year}\right)^{1/2} {\text{\AA}}.
\end{equation}
As can be seen, the transition wavelength lies at UV/optical bands for an SMBHB with  an orbital period of several years and a total mass of $10^8M_\odot$ accreting at 0.1 times the Eddington accretion rate. In Figure~\ref{fig:lambdac}, by setting $f=1$, we plot $\lambda_{\rm tran}$ with $M_{\rm t}$, $\dot m_{\rm c}$, and $a_{\rm B}$ using Equation~(\ref{eqn_lam_break}). Note that the above two equations should be regarded as only a rough approximation. The real transition wavelength should be calculated from the $\tau-\lambda$ relation as described in Section~\ref{sec_relation}, which might slightly differ from the above approximation.

\section{Application to the SMBHB candidate PG~1302-102}\label{sec_pg1302}
PG~1302-102 was identified as an SMBHB candidate by \cite{Graham2015} based on the evidence for a periodicity in its long-term optical variations. The period was favored by subsequent studies, including \cite{Kovacevic2019} and \cite{Zhu2020}, although one still cannot rule out the possibibity that the periodicity is a false positive caused by the red-noise variability of AGNs (\citealt{Vaughan2016}). The optical spectrum of PG 1302-102 shows asymmetric broad emission lines, probably indicative of a disturbed broad-line region surrounding the putative SMBHB (e.g., \citealt{Rigamonti2025}). Recently, \citet{Liu2024} conducted an intensive multiwavelength monitoring campaign of PG 1302-102 with the {\it Swift} and the Las Cumbres Observatory (LCO) network telescopes during the period MJD 59551-59817. The campaign tested the continuum RM signature of a simple mini-disk model and found that the estimated disk size is consistent within uncertainties with the theoretical anticipation for a single SMBH disk. Nevertheless, their interband time lags measured with the ICCF method showed marginal discrepancies (although not significant) between the {\it Swift} and LCO observations, which were ascribed to the different sampling patterns. 
Building on the dataset of \cite{Liu2024}, we have merged the {\it Swift} and LCO light curves in similar bands to alleivate the issue of the sampling patterns.
We then re-measure the inter-band time lags using two methods, one is the ICCF and the other is the Bayesian method \texttt{MICA}\footnote{\url{https://github.com/LiyrAstroph/MICA2}} (\citealt{Li2016MICA}). We finally apply the SMBHB model to fit the inter-band time lags and make a comparison with the single SMBH model. Throughout the analysis, we use the redshift $z=0.2784$ for PG 1302-102.

\subsection{Measuring the inter-band time lags}\label{sec:measuring lags}

\begin{figure*}
\centering
    \includegraphics[width=0.85\textwidth]{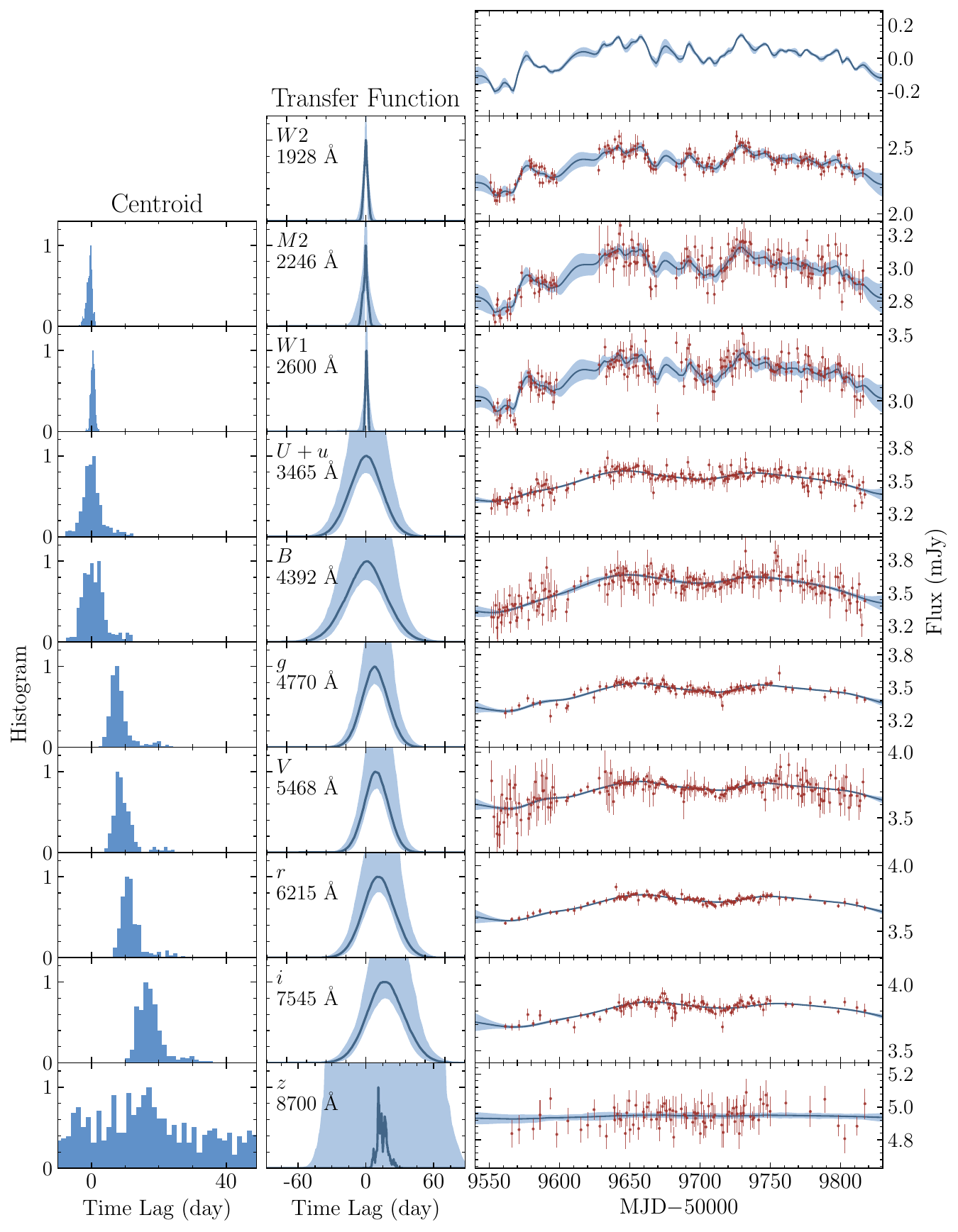}
    \caption{The inter-band time lag analysis of PG~1302-102 using \texttt{MICA}. (Left) Histograms of MCMC samples of the centroid time lags (observed-frame) of the Gaussian transfer functions. (Middle) The inferred Gaussian transfer functions (solid lines) and $1\sigma$ uncertainties (shaded bands). (Right) The observed light curves and reconstructions with 1$\sigma$ uncertainties. The topmost panel shows the virtual driving light curve.}
    \label{fig:mica}
\end{figure*}

We merge the datasets of the {\it Swift} $U$- and LCO $u$-band, {\it Swift} $B$- and LCO $B$-band, and {\it Swift} $V$- and LCO $V$-band, respectively. 
The Bayesian package \texttt{PyCALI}\footnote{\url{https://github.com/LiyrAstroph/PyCALI}}~\citep{Li2014} is used to do merging. Specifically, the light curve is scaled and shifted with a multiplicative factor $\varphi$ and an additive factor $G$ as $F_{\rm cali} = \varphi F_{\rm obs} -G$. The best values of $\varphi$ and $G$ are determined by the Markov-chain Monte Carlo (MCMC) technique (see \citealt{Li2014}). 
After merging, we have 10 bands of light curves, as shown in the rightmost panels of Figure~\ref{fig:mica}. 
Using the best reconstructed light curves from \texttt{PyCALI}, we identify outlier points as those with the deviations from the reconstruction more than 3$\sigma$ and discard them in subsequent time-lag analysis.

We then employ the ICCF and \texttt{MICA} methods to measure the inter-band time lags. For the ICCF method, the $W2$ band is set as the reference band. Following the convention, the inter-band time lag is determined by the centroid of the CCF above 80\% of the peak value. The uncertainties are estimated by the 68.3\% percentiles of the cross-correlation centroid distributions (CCCDs), generated from Monte Carlo simulations with the flux randomization and random subset selection (FR/RSS) method (\citealt{Peterson1998}).

\texttt{MICA} is an alternative method that is not based on the CCF. Instead, it describes the driving light curve using the damped random walk (DRW) process and models the inter-band transfer function with some basic functions (such as Gaussian, top hat, and exponetial; \citealt{Li2016MICA}). The parameters of the transfer function are sampled by the MCMC technique with the diffusive nested sampling alogrithm (\citealt{Brewer2010}). 
For simplicity, we adopt one Gaussian for the transfer function, with the Gaussian amplitude, center, and standard deviation as free parameters. The prior of the Gaussian center is set to be uniform over the range (-10, 50) days and the prior of the Gaussian standard deviation is set to be log-uniform over (1, 40) days.
The inter-band time lag is assigned as the center of the Gaussian and the uncertainties are assigned as the 68.3\% percentiles of the posterior distribution of the Gaussian center.

It is worth mentioning that the ICCF time lags slightly depend on the choice of the reference band. This is because the ICCF method relies on the interpolation of light curves and a different reference light curve will induce different interpolated variation patterns. For \texttt{MICA}, we use the \texttt{vmap} mode, which creates a virtual driving light curve and models all bands of light curves with respect to this virtual light curve simultaneously. The virtual driving light curve is set to have a zero lag with respect to the $W2$ band. As such, light curves in all bands have similar weights in the time-lag analysis and the issue of adopting the reference band in the ICCF method is largely alleivated. 

Below, we denote the time lag from ICCF and \texttt{MICA} by $\tau_{\rm ICCF}$ and $\tau_{\rm MICA}$, respectively, and list their values in Table~\ref{tab:best-fit-tau}. As observed, these two kinds of time lags are generally consistent with each other within uncertainties, except that the MICA time lags typically have smaller uncertainties. The reasons are two-fold. First, the \texttt{MICA} method presumes the form of the transfer function and models all the light curves simultaneously. The latter makes the \texttt{MICA} results less sensitive to the sampling gaps. Second, as demonstrated by \cite{Peterson1998}, the FR/RSS procedure used to estimate the uncertainties in the ICCF method  quotes conservative uncertainties, usually larger than the real values. We show the \texttt{MICA} time lags with wavelengths in Figure~\ref{fig:1302fit} and the ICCF time lags in Appendix~\ref{app:ICCF}. A transition feature around 3000-4000 {\AA} can be seen in the \texttt{MICA} time lags, despite it being not signficant. Such a transition is not obvious in the ICCF time lags, probably because of the large uncertainties.
 
\begin{table}
    \centering
    \caption{The inter-band time lags with respect to $W2$ band from MICA and ICCF. Both the wavelength $\lambda_{\rm 0}$ and time lag $\tau$ are given in the observed-frame.}
    \label{tab:best-fit-tau}
    \begin{tabular}{cccc} 
		\hline
		\specialrule{0em}{1pt}{1pt}
		~~~~~~Band~~~~~~ & $\lambda_0 $ & ~~~~~~~~$\tau_{\rm MICA}$~~~~~~~~ & $\tau_{\rm ICCF}$  \\
		& (\AA) & (day) & (day) \\\hline
		\specialrule{0em}{2pt}{2pt}
		$W2$ & 1928 & $0$ & $0.00_{-0.98}^{+0.97} $\\
		\specialrule{0em}{2pt}{2pt}
		$M2$ & 2246 & $-0.33_{-1.02}^{+0.63} $ & $-0.49_{-1.56}^{+1.62} $\\
		\specialrule{0em}{2pt}{2pt}
		$W1$ & 2600 &$ 0.54_{-0.65}^{+0.53} $ & $0.06_{-1.88}^{+1.89} $\\
		\specialrule{0em}{2pt}{2pt}
        $U+u$ & 3465 &$0.65_{-2.60}^{+2.56} $ & $2.01_{-2.83}^{+2.80} $\\
        \specialrule{0em}{2pt}{2pt}
		$B$ & 4392 &$0.96_{-2.83}^{+3.41}$ & $3.53_{-3.12}^{+3.45} $\\
		\specialrule{0em}{2pt}{2pt}
		$g$ & 4770 &$8.38_{-1.99}^{+2.80}$ & $10.58_{-4.11}^{+5.88} $\\
		\specialrule{0em}{2pt}{2pt}
		$V$ & 5468 &$9.46_{-1.72}^{+2.82}$ & $7.98_{-5.00}^{+5.10} $\\
		\specialrule{0em}{2pt}{2pt}
		$r$ & 6215 &$11.77_{-2.00}^{+3.13}$ & $13.95_{-3.82}^{+5.11} $\\
		\specialrule{0em}{2pt}{2pt}
		$i$ & 7545 &$17.66_{-2.74}^{+4.26}$ & $18.61_{-7.35}^{+7.91} $\\
		\specialrule{0em}{2pt}{2pt}
		$z$ & 8700 &$15.85_{-15.30}^{+19.97}$ & $21.65_{-15.68}^{+24.37} $\\
		\specialrule{0em}{2pt}{2pt}
		\hline
    \end{tabular}
\end{table}

\begin{figure*}
\centering
    \includegraphics[width=0.7\textwidth]{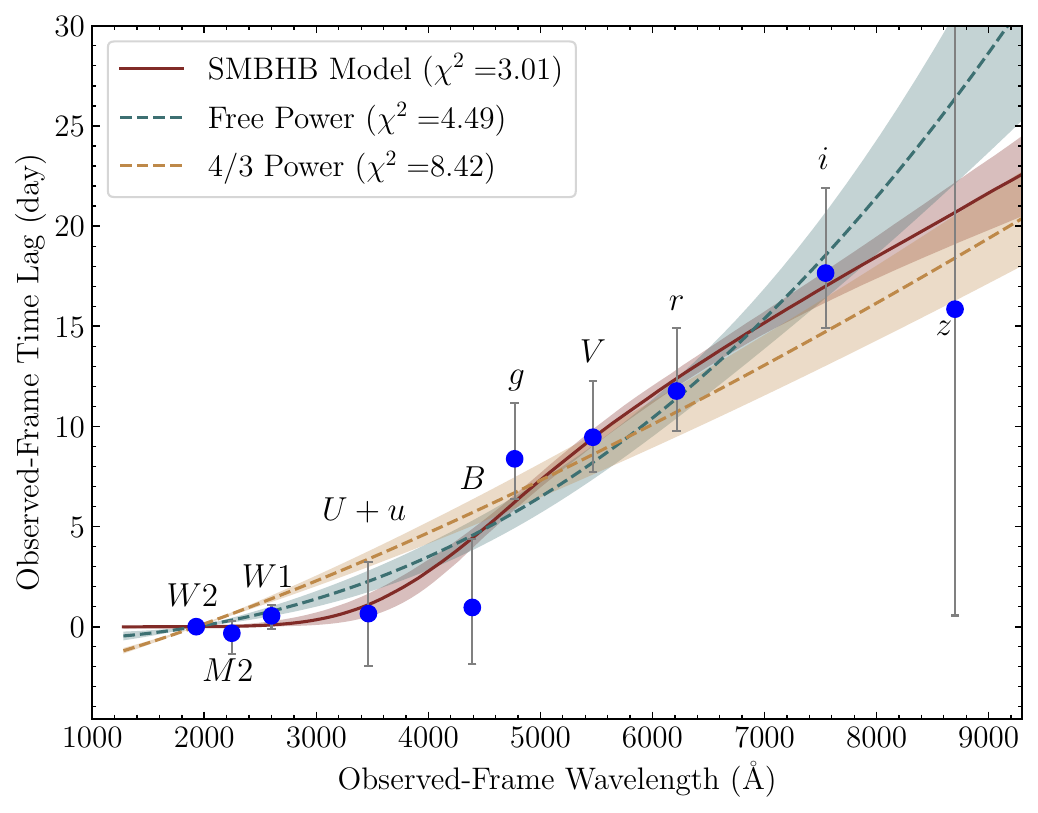}
    \caption{The $\tau(\lambda)$ relation of PG~1302-102 and the fits using the SMBHB model. The blue points represent the inter-band time lags measured from \texttt{MICA}. The red solid line represents the central lags of the 1000 best-fitting parameter groups selected from the posterior samples.
    The yellow and green dashed lines represent the best fits using the power law $\tau_{\rm obs} = \tau_0(1+z) [(\lambda/\lambda_0)^{\beta}-1]$ with $\beta = 4/3$ and set to be free, respectively.  Here, $\lambda_0 = 1928~\angstrom$ corresponds to the central wavelength of the $W2$ band (corresponding to 1508~{\angstrom} in the rest frame). Shaded bands represent the 1$\sigma$ uncertainties.
    }
    \label{fig:1302fit}
\end{figure*}

\begin{figure*}
\centering
    \includegraphics[width=0.95\textwidth]{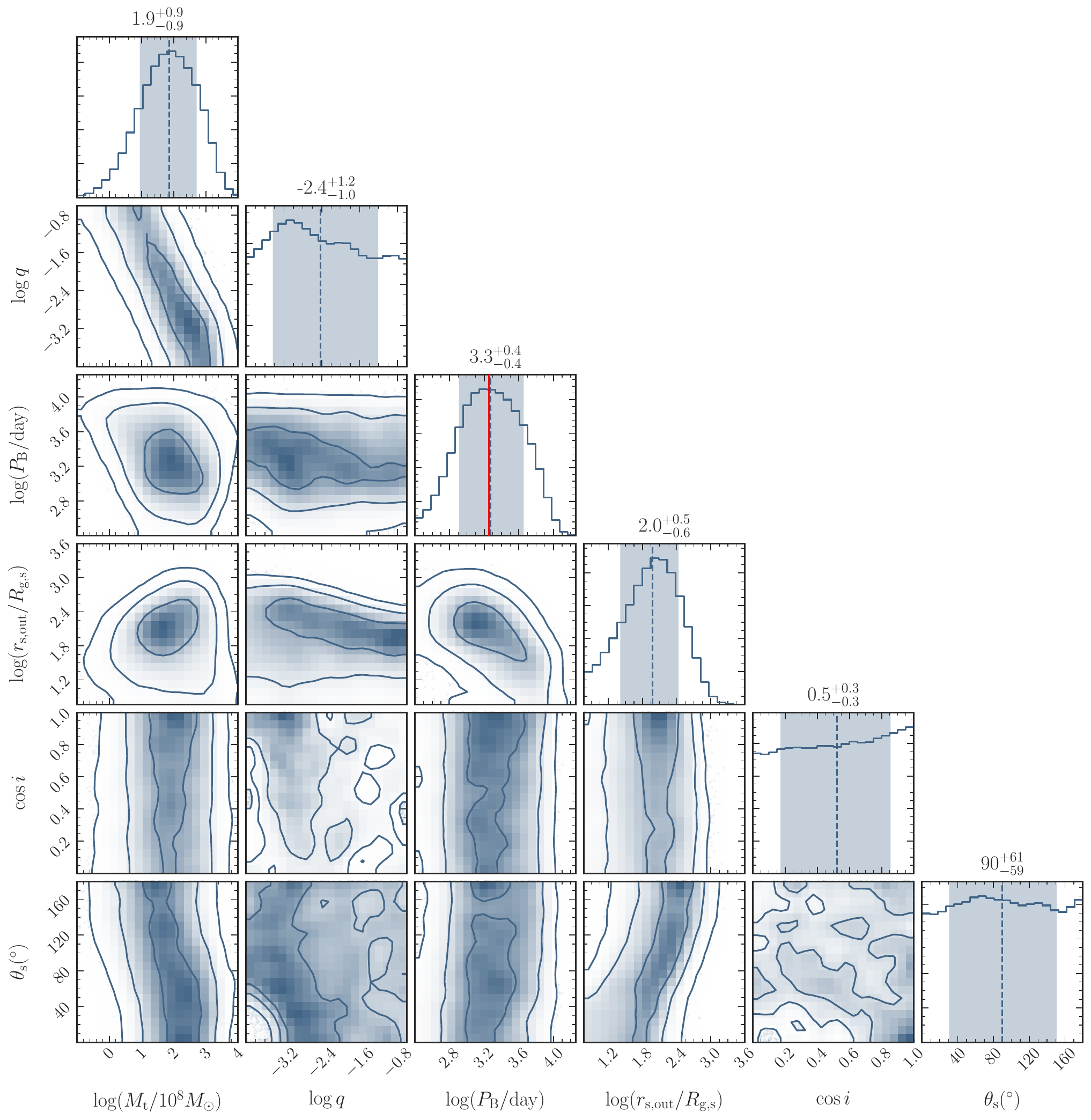}
    \caption{The posterior distribution of the orbital parameters obtained by fitting the \texttt{MICA} time lags. In the diagonal panels, the vertical dashed line represents the median value and the shaded band represents the 68.3\% confidence interval. The vertical red solid line marks $\log (P_{\rm B}/{\rm day})=3.25$. The contours are at $1\sigma$, $2\sigma$, and $3\sigma$ levels.
    }
    \label{fig:corner}
\end{figure*}

\begin{table*}
    \centering
    \caption{The free orbital parameters, priors, and the best-fit values using the ICCF and \texttt{MICA} time lags in two cases, one with the orbital period $P_{\rm B}$ set to be free and the other with fixed period  $\log(P_{\rm B}/{\rm day})=3.25$, corresponding to 1796 days in the rest frame. For the free case, the lower limit of $P_{\rm B}$ is set to be the duration of the light-curve data and the upper limit is set to be about 50 yr, corresponding to  $\log(P_{\rm B}/{\rm day})\approx4.3$. The outer edge of the mini-disk $r_{\rm s, out}$ is additionally limited by the condition that it shoud be smaller than the Roche-lobe size, i.e., $r_{\rm s, out}<r_{\rm R, s}$. }
    \label{tab:best-fit-para}
    \begin{tabular}{lcccccc} 
		\hline
		~~~~~~~~~~~Parameter~~~~~~~~~~~ & ~~~~~~~~~~~~Prior~~~~~~~~~~~~ &\multicolumn{2}{c}{~~~~~~~~~~~Value ($P_{\rm B}$ free) ~~~~~~~~~~~} & & \multicolumn{2}{c}{Value ($P_{\rm B}$ fixed)}\\\cline{3-4}\cline{6-7}
		~~~&~~~&~~~MICA~~~&~~~ICCF~~~& &~~~MICA~~~&~~~ICCF~~~\\
		\hline
		\specialrule{0em}{2pt}{2pt}
		$\log( M_{\rm t}/10^8 M_{\odot})$ & Uniform[$-1.0$, $4.0$] 
		& $1.9^{+0.9}_{-0.9}$  & $1.8^{+0.9}_{-1.0}$ & 
		& $1.9^{+0.8}_{-0.9}$  & $1.9^{+0.9}_{-1.0}$   \\
		\specialrule{0em}{2pt}{2pt}
		$\log q$ & Uniform[$-4.0$, $\log(0.25)$] 
		& $-2.4^{+1.2}_{-1.0}$ & $-2.4^{+1.1}_{-1.0}$ & 
		& $-2.4^{+1.2}_{-1.1}$ & $-2.3^{+1.1}_{-1.0}$ \\
		\specialrule{0em}{2pt}{2pt}
		$\log (P_{\rm B}/{\rm day})$ & Uniform[$2.4$, $4.3$] 
		& $3.3^{+0.4}_{-0.4}$ & $3.2^{+0.4}_{-0.4}$& 
		& 3.25 & 3.25 \\
		\specialrule{0em}{2pt}{2pt}
        $\log (r_{\rm s, out}/R_{\rm g,s})$ & Uniform[$\log(6.0)$, $4.0$] 
        & $2.0^{+0.5}_{-0.6}$ & $1.9^{+0.5}_{-0.6}$ & 
        & $2.1^{+0.4}_{-0.5}$ & $2.0^{+0.5}_{-0.5}$ \\
        \specialrule{0em}{2pt}{2pt}
		$\cos i$ & Uniform[0, 1] 
		& $0.5^{+0.3}_{-0.3}$ & $0.5^{+0.3}_{-0.3}$ & 
		& $0.5^{+0.3}_{-0.4}$ & $0.5^{+0.3}_{-0.3}$ \\
		\specialrule{0em}{2pt}{2pt}
		$\theta_{\rm s}/{\rm degree}$ & Uniform[0, 180] 
		& $95^{+61}_{-59}$ & $94^{+60}_{-60}$ & 
		& $85^{+64}_{-57}$ & $92^{+60}_{-61}$ \\
 		\specialrule{0em}{2pt}{2pt}
		\hline
    \end{tabular}
\end{table*}

\begin{figure*}
\centering
    \includegraphics[width=0.8\textwidth]{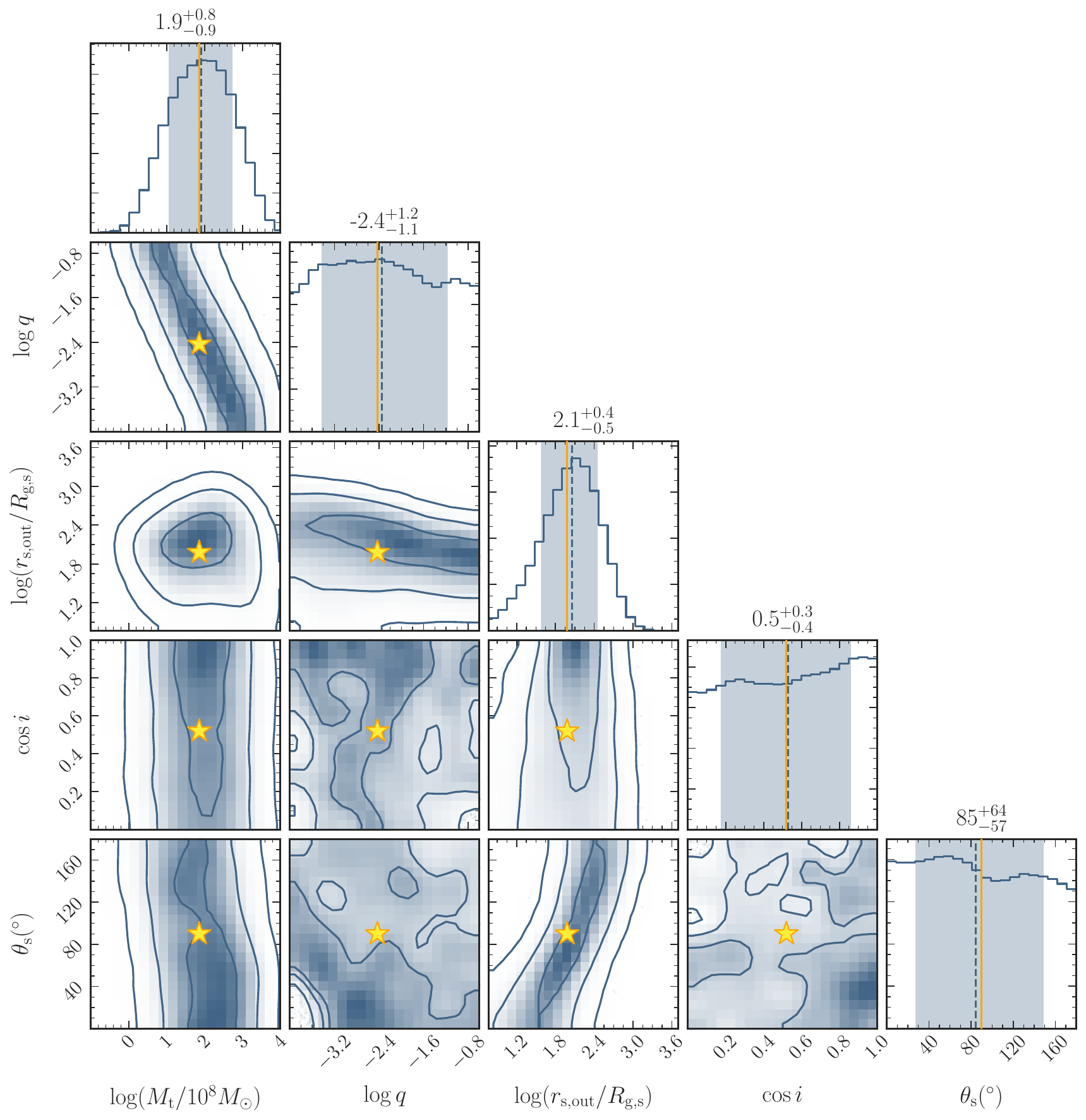}
    \caption{Same as Figure~\ref{fig:corner}, but with the orbital period $P_{\rm B}$ fixed to 1796 days. The orange solid lines and star markers indicate the median values of the posterior distribution from Figure~\ref{fig:corner}. The contours are at $1\sigma$, $2\sigma$, and $3\sigma$ levels.
    }
    \label{fig:corner2}
\end{figure*}

\subsection{Fitting the inter-band time lags with the SMBHB model}\label{sec:fitting}
As illustrated above, the parameters affecting the inter-band time lags mainly include the total mass $M_{\rm t}$, the mass ratio $q$, the orbital period $P_{\rm B}$ (or $a_{\rm B}$), $r_{\rm s, out}$, the inclination $i$, and the orbital phase angle $\theta_{\rm s}$. The accretion rates $\dot M_{\rm c}$ and $\dot M_{\rm s}$ are also crucial but can be estimated from the observed luminosity. The remaining parameters, $L_{\rm b}$, $h_{\rm s}$, $r_{\rm s, in}$ and $r_{\rm c, out}$, are fixed at the fiducial values listed in Table~\ref{tab:para}.  We adopt a steady total bolometric luminosity of $L_{\rm bol}=6.5\times10^{46}~{\rm erg~s^{-1}}$ for PG 1302-102, resulting in an accretion rate of $\dot M_{\rm c}=L_{\rm bol}/\eta c^2=3.7~M_\odot~{\rm yr^{-1}}$ with the radiative efficiency $\eta=0.3$ (\citealt{D'Orazio2015}). Here, the bolometric luminosity is estimated from the $V$-band luminosity with a bolometric correction factor of 10 (\citealt{D'Orazio2015}). This estimate neglects the differences in the spectral energy distribution between SMBHBs and single SMBHs and assumes that a common bolometric correction factor is applicable. We note from Equation~(\ref{T}) that the accretion rate is indeed degenerate with the total mass $M_{\rm t}$ so that any uncertainty in the adopted bolometric luminosity and radiative efficiency will be finally transferred to the inferred total mass.

Given a set of model parameters, denoted by $\bm{\Theta}$, the $\tau(\lambda)$ relation can be computed following the method in Section~\ref{sec_method}. Note that, to compare with the observed time lags, we need to additionally subtract the theoretical time lag at $W2$ band from all the calculated $\tau$. We then employ the Bayesian method to determine the parameter values. The posterior probability is given by 
\begin{equation}\label{eqn_posterior}
    P(\bm{\Theta}|\bm{\tau}_{\rm obs})\propto P(\bm{\Theta})e^{-\chi^2/2},
\end{equation}
where $P(\bm{\Theta})$ denotes the parameter priors, $\bm{\tau}_{\rm obs}$ detnotes the observed time lags, and $\chi^2$ is the chi square calculated as
\begin{equation}\label{eqn_chi2}
\chi^2 = \sum_i \frac{(\tau_{{\rm obs}, i}-\tau_i)^2}{\sigma^2_{{\rm obs}, i}},
\end{equation}
where $\tau_{{\rm obs}, i}$ and $\sigma_{{\rm obs}, i}$ denote the observed time lag and uncertainty at the $i$-th band, respectively, and $\tau_i$ denotes the theoretical time lag.
For the parameters $M_{\rm t}$, $P_{\rm B}$, and $r_{\rm s,out}$, which are expected to span a large range, a log-uniform prior is used, while for $\cos i$ and $\theta_{\rm s}$, a uniform prior is used. The prior ranges are listed in Table~\ref{tab:best-fit-para}. The observed time lags usually have asymmetric lower and upper errors. We simply assign $\sigma_{{\rm obs}, i}$ in Equation~\eqref{eqn_chi2} as the mean of the lower and upper errors. We employ the package \texttt{emcee}\footnote{\url{https://github.com/dfm/emcee}} (\citealt{emcee}) to explore the posterior probability and generate the posterior samples, from which we determine the best estimates and uncertainties of the model parameters as the meidans and 68.3\% confidence intervals, respectively. We run \texttt{emcee} for a total of $6 \times 10^4$ steps, with the first $10^4$ steps discarded as burn-in.

Table~\ref{tab:best-fit-para} summarizes the inferred values and uncertainties of the model parameters. We find a good consistency within uncertainties between the results from the ICCF and MICA time lags. Below, we present the fitting results using the MICA time lags and make a comparison with those from the ICCF time lags in Appendix~\ref{app:ICCF}. 

We show the best fits of the SMBHB model to the MICA time lags in Figure~\ref{fig:1302fit} and the obtained posterior distributions of the model paramters in Figure~\ref{fig:corner}. Among the six free parameters, the total mass $M_{\rm t}$, the orbital period $P_{\rm B}$, and the outer radius of the mini-disk $r_{\rm s, out}$ show significant posterior peaks, while the three remaining parameters (the mass ratio $q$, inclination $i$, and orbital phase angle $\theta_{\rm s}$) are poorly constrained. There is a strong anti-correlation between $M_{\rm t}$ and $q$. This is because $M_{\rm t}$ and $q$ jointly determine the mass of the secondary SMBH and thereby control the time lags at short wavelengths ($W2$, $M2$, $W1$, and $U+u$). 
It is interesting to note that the inferred total mass $\log(M_{\rm t}/M_\odot)=9.9_{-0.9}^{+0.8}$ is consistent within uncertainties with the independent estimate using the broad emission lines, which lies in a range of $\log(M_\bullet/M_\odot)=8.3-9.4$ (\citealt{Graham2015}). The inferred outer edge of the mini-disk is $r_{\rm s, out}\approx 100\, R_{\rm g, s}$, much smaller than the Roch-lobe size ($\approx 1000\, R_{\rm g, s}$, corresponding to $\xi\approx0.1$ in Equation~\ref{eqn_r_outs}), indicating that the mini-disk is significantly truncated (e.g., \citealt{Artymowicz1994}). In addition, the inferred orbital period $\log(P_{\rm B}/{\rm day})= 3.3_{-0.4}^{+0.4}$ also approximately matches with the observed period of $\log(P/{\rm day})\sim 3.25$ in the long-term optical variations of PG 1302-102 (see Appendix~\ref{app:period}). This is remarkable considering that the present SMBHB fitting is completely independent from the periodicity search.

If using a total mass of $\log(M_{\rm t}/M_\odot)=9.87$, the mass ratio of $\log q = -2.45$, and the mass accretion rate $\dot M_{\rm c}=\dot M_{\rm s}=3.7~M_\odot~{\rm yr}^{-1}$,  the dimensionless accretion rate for the circumbinary disk is $\dot m_{\rm c}=0.022$ and for the mini-disk is $\dot m_{\rm s}=6.28$. This indicates that the secondary SMBH is accreting at a super-Eddington rate. In Appendix~\ref{app:sed}, we caclulate the spectral energy distributions (SEDs) of the mini-disk and circumbinary disk using the parameters selected from the above posterior sample and make a comparison with the observations of PG 1302-102. Considering the long-term variations of PG 1302-102 ($\sim0.4$ mag in $V$ band; see Appendix~\ref{app:period}), we find that the observed SED is reasonably well reproduced (except for the data redward of 1.5~$\mu$m, which might be dominated by emissions from the outer torus), implying consistency between our SMBHB fitting results and SED data.

To compare with the single SMBH model, in Figure~\ref{fig:1302fit}, we also superimpose the fits using a power law as
\begin{equation}\label{eqn_powerlaw}
\tau_{\rm obs} = \tau_0 (1+z)\left[\left(\frac{\lambda}{\lambda_0}\right)^{\beta}-1\right],
\end{equation}
where $\lambda_0 = 1928~\angstrom$ is the central wavelength of the $W2$ band (corresponding to 1508~{\angstrom} in the rest frame) and $\tau_0$ and $\beta$ are free parameters, which denote the rest-frame time lag at $\lambda_0$ and the power-law index, respectively. We consider two cases, one is setting $\beta$ free and the other is fixing $\beta$ as the aniticpated value of the standard accretion disk, namely, $\beta=4/3$. The former case yields $\tau_0 = 0.58 \pm 0.34$ days and $\beta = 2.39 \pm 0.46$ and the latter case yields $\tau_0 = 2.23 \pm 0.26$ days.

The SMBHB model and the two cases of power-law fitting give a minimized $\chi^2$ of $\chi^2_{\rm min}=$3.01, 4.49 ($\beta$ free), and 8.42 ($\beta$ fixed), respectively. If using the Bayesian information criterion (BIC\footnote{BIC is defined as ${\rm BIC}=k\ln n + \chi_{\rm min}^2$, where $k$ is the number of model parameters and $n=9$ is the number of data points. For the SMBHB model, since the three parameters ($q$, $i$, and $\theta_{\rm s}$) are poorly constrained, we use $k=3$.}; \citealt{Schwarz1978}), the three models yield BIC values of 9.6, 8.9, 10.6, respectively, suggesting that the models are comparable in terms of goodness of fit. Nevertheless, for the case of free $\beta$, the obtained $\beta$ significantly deviates from the canonical value of $4/3$ by more than the 5$\sigma$ confindence interval. Moreover, the obtained $\tau_0$ is unrealistically small, compared to the anticipated value of $2.30$ days from the standard accretion disk model. 

PG 1302-102 displays sinusoid-like periodical variations in its long-term optical light curve (\citealt{Graham2015, Kovacevic2019, Zhu2020}). If we assume that this periodicity arises from the orbital motion of the SMBHB, the observed period provides a strong prior on our SMBHB model. In Appendix~\ref{app:period}, we compile the latest archival photometric data and merge various sources of $g$ and $V$ photometry to obtain an optical light curve between MJD 52668 and 60538, spanning a temporal baseline of 7870 days. It seems that both the sinusoidal period and amplitde are gradually increasing by a simple visual inspection. The cause for such changes is unknown, probably arising from additional variations superimposed on top of the periodic signal (e.g., \citealt{Kovacevic2019}). Nevertheless, if still using a sinusoid to fit the whole light curve, we obtain a rest-frame period of $1796$ days (see Appendix~\ref{app:period} for the detail), which, as expected, is slightly longer than the rest-frame period of $\sim$1474 days reported by \cite{Graham2015}. We assume that the period represents the orbital period of the SMBHB and redo the SMBHB fitting by fixing $P_{\rm B}=1796$ days (corresponding to 3.25 in logarithmic scale). The obtained parameter values are listed in Table~\ref{tab:best-fit-para} and the posterior distributions are plotted in Figure~\ref{fig:corner2}. As expected, the inferred values of all free parameters are consistent with those with free $P_{\rm B}$, except for the slightly smaller uncertainties in $M_{\rm t}$ and $r_{\rm s, out}$.

\section{Discussions}\label{sec_discussions}
\subsection{Limitations of the SMBHB model and future improvements}\label{sec:limitations}
Our SMBHB model makes several assumptions to simplify calculations. Below, we briefly discuss these assumptions and give remarks on future improvements. 

One major assumption is that the reverberation of the accretion disks around SMBHBs follows the popular lamp-post scenario (e.g., \citealt{Cackett2021}). This would imply a correlation between X-ray and UV/optical variability. However, in observations of normal AGNs, the correlation between X-ray and UV/optical is diverse: some show such a weak or sometimes strong correlation but some show no correlation (\citealt{Starkey2017, Cackett2018, Cackett2020, Edelson2019, Edelson2024, GonzalezBuitrago2022, Kara2021, Kara2023}). This imposes a challenge for the simple lamp-post model and more complicated process might be taking place (e.g., \citealt{Gardner2017, Sun2020, Kara2023, Antonucci2023, Yaqoob2023}). Despite this challenge, strong correlations between UV and optical variability are generally observed. As expected, the observed inter-band time lag also increases with wavelength. In this sense, the basic reverberation picture for accretion disks is qualitatively valid. Therefore, regardless of the underlying processes, the reverberation principle can be phenomenologically applied to the accretion disks surrounding SMBHBs. The presence of a cavity with low gas density will naturally cause null/weak responses of the accretion disks at some specific wavelengths, resulting in a transition feature in the $\tau(\lambda)$ relation.

Another major assumption is that accretion from the circumbinary disk occurs predominately onto the secondary SMBH so that the radiation from the mini-disk surrounding the primary SMBH is negligible. This assumption is valid for low-mass-ratio SMBHBs and well supported by numerical simulations (\citealt{Farris2014, Shi&Krolik2015, Bowen2019, Combi&Lopez2022}). As the mass ratio of the SMBHB increases, the radiation from the primary SMBH becomes no longer negligible. In this case, each SMBH carries a mini-disk associated with a corona. In other words, there are two separate sources illuminating the accretion disks, driving complicated variability in the multi-band light curves. Moreover, because of the presence of two driving sources, the simple time lags from the ICCF method are likely to be dominated by whichever SMBH exhibits  stronger variations, with the other SMBH causing bias and/or noise due to slow and/or fast variations from the sub-dominant SMBH. More sophisticated analysis methods need to be developed to decouple the mixing variability in a wavelength band and determine the time lags/transfer functions associated with each driving source.

In calculating transfer functions and time lags, we implicitly assume that the orbital phase $\theta_{\rm s}$ remains unchanged.  In practice, the secondary SMBH is in orbital motion and the transfer functions are indeed time dependent. Our neglect of changes in $\theta_s$ is justified when the orbital period is much longer than the duration of RM monitoring. The orbital phase angle $\theta_{\rm s}$ does not affect reverberation of the mini-disk because it moves along with the secondary SMBH. However, $\theta_{\rm s}$ does affect the time lags at long wavelengths where the emissions mainly stem from the circumbinary disk (see Section~\ref{sec:dependence}), particularly when the inclination is large (more edge-on). It is easy to show that for a face-on inclination ($i=0$), the transfer functions and time lags are independent of $\theta_{\rm s}$. For a moderate inclination, Figure~\ref{fig:lam-tau} illustrates that $\theta_{\rm s}$ can induce differences in time lags by few times as it changes from 0 to $180^\circ$. Nevertheless, the transition feature in the $\tau(\lambda)$ relation is generally retained. For the cases of short orbital period, it is worth investigating the influences of time-dependent transfer functions on the time-lag measurement and developing a sophisticated analysis to subtract the SMBHB orbital information in a future work. 

We simply consider the shape of the cavity as circular, centered at the SMBHB's barycenter. This is a quite rough approximation. Previous numerical simulations illustrated that the cavity might be irregular in shape and evolve with time.  Tidal streams, pulled off the inner edge of the circumbinary disk, penetrate through the cavity and feed the mini-disks (e.g., \citealt{DOrazio2016, Combi&Lopez2022, Westernacher2022, Whitley2024}). These tidal streams move on nearly ballistic orbits and hit the mini-disk at supersonic speeds, generating shocks and X-ray excess emissions (e.g., \citealt{Roedig2014, Ryan2017}). Moreover, spiral arms might be excited by the tidal torques of the SMBHB in both the mini-disk and circumbinary disk, leading to extra variations in the emission fluxes. All these processes complicate the present simple reverberation model. Nevertheless, the basic principle that the presence of a cavity in SMBHBs causes their continuum reverberation to be distinctive from those of single SMBHs remains valid.

Finally, an improvement can be made to the present SMBHB model is that, instead of calculating the centroid time lags, we can directly apply the transfer functions to fit the observed multi-band light curves, an approach similar to the disk reverberation analysis for single SMBHs developed by \cite{Starkey2016}. As such, we can alleviate the degenercy between the model parameters arising from solely fitting the $\tau(\lambda)$ relation (see Section~\ref{sec_pg1302}) and obtain more stringent contraints on the model parameters.

\subsection{Implications for SMBHB search}
Multi-band, high-cadence time-domain surveys provide ideal datasets to apply the present SMBHB models. Such surveys potentially include (but are not limited to) the Legacy Survey of Space and Time (LSST; \citealt{Ivezic2019}) conducted with the Vera C. Rubin Observatory in the Southern Hemisphere and the Wide Field Survey (WFS) conducted with the 2.5m telescope of the University of Science and Technology of China and the Purple Mountain Observatory in the Northern Hemisphere (\citealt{Wang2023}). The LSST survey can deliver photometric light curves data in six bands ($ugrizy$) at a cadence of every several days. The WFS survey can deliver photometric light curve data in four bands ($ugri$) at a daily cadence. From these high-cadence light curves, a feasible search strategy is as follows. Firstly, one can efficiently estimate the inter-band time lags using the ICCF method and obtain the $\tau(\lambda)$ relations in each monitoring season. Considering the limited wavelength coverage of the available bands, it might be difficult to observe the complete transition feature in the the $\tau(\lambda)$ relations. Those sources with exceptional slopes in the $\tau(\lambda)$ relations, significally smaller or larger than the canonical 4/3, can be selected as potential SMBHB candidates. Follow-up detailed analysis of those candidates with more sophisticated methods (such as MICA) are then employed to make a further verification.   

According to Equation~(\ref{eqn_lam_break_p}), the transition wavelengths at UV/optical bands typically occurs for more massive SMBHs ($>10^8M_\odot$) with relatively shorter orbital periods (several years). These SMBHB candidates might constitute the major sources of the nano-Hertz gravitational wave radiation background and are the best observation targets of pulsar timing arrays. 

\section{Conclusions}\label{sec_conclusions}

We develop a new approach for identifying low-mass-ratio SMBHBs in active galactic nuclei through multi-band continuum reverberation mapping, based on the principle that the low-density cavity in the accretion disks of SMBHBs naturally gives rise to distinctive continuum reverberation features compared to that of single SMBHs. The cavity with a size comparable to the SMBHB's orbital separation is created by the SMBHB's tidal torque and has been extensively confirmed in numerical simulations. The formation of the cavity induces a radiation deficiency of the accretion disks and thereby reduce the associated disk responses at certain wavelengths. As a result, the relation between the inter-band time lag and wavelength ($\tau-\lambda$) deviates from the the canonical $\tau \propto \lambda^{4/3}$ power law observed in single SMBH systems. Specifically, the $\tau-\lambda$ relation appears flat at shorter wavelengths becasue of the truncated sizes of the mini-disks. It transitions to the $\lambda^{4/3}$ power law at longer wavelengths where the radiation is dominated by the circumbinary disk.

Following the popular lamp-post scenario and assuming that only the secondary SMBH is active in a low-mass-ratio system, we devise a simple continuum reverberation model to calculate revererbation signatures of the accretion disks surrounding SMBHBs. In this model, a hot corona located above the secondary black hole emits variable X-ray photons that illuminate both the mini-disk and the circumbinary disk, which are described by the standard thin disk model. A fraction of the incident X-ray photons are absorbed and reprocessed into UV/optical emissions, leading to responses to the variable X-ray irradiation with time lags due to the light travel time. We calculate the transfer functions and explore the dependence of the $\tau-\lambda$ relation on the SMBHB's orbital parameters. The tranfer functions typically exhibit a bimodal structure, separated by a null region owing to the cavity. The resulting transition feature in the $\tau-\lambda$ relation occurs at the wavelength approximately assoicated with the characteristic temperature at the inner edge of the circumbinary disk. For an SMBHB with a total mass of $10^8M_\odot$ and an orbtial period of several years, the transition wavelength approximately lies at UV/optical bands.

We apply the SMBHB model to the intensive multi-wavelength monitoring data of the SMBHB candidate PG 1302-102 (\citealt{Liu2024}) identified from the long-term periodical variations in optical (\citealt{Graham2015}). We measure the inter-band time lags using the Bayesian method \texttt{MICA} and tentatively observe a potential transition around 3000~{\AA}-4000~{\AA} (in the observed frame) although it is not statistically significant. We use the SMBHB model to fit the inter-band time lags and derive constraints on the key parameters including the total mass, orbital period, and the outer edge of the mini-disk. The SMBHB model provides a goodness of fit to the inter-band time lags comparable to the single power-law model expected from disk reverberation of single SMBHs. Nevertheless, the inferred orbital period shows a remarkable agreement with the period derived from the long-term optical variability, plausibly implying that the SMBHB model is reasonable. With the current data quality of PG~1302-102, we cannot discriminate between the SMBHB model and the power-law models. Future higher-quality continuum revereberation mapping data (with an improved sampling rate and reduced measurement uncertainties) will be essential for a firm conclusion.

It is worth mentioning that the transition feature can also be produced by a bowl-like disk model without invoking of SMBHBs (\citealt{Starkey2023, Edelson2024}). Unlike the traditional assumption that the disk thickness is negligible, this model instead incorporates a finite, radial-dependent thickness profile. The light travel lag and the illumination incidence angle of the lamp-post on the disk surface are changed and thereby the resulting $\tau-\lambda$ relation can also deviate from the conocial $\lambda^{4/3}$ power law, depending on the adopted form of the thickness profile. It is expected that the transfer functions of bowl-like disks are still different in shape from the bimodal transfer functions of the SMBHB model. Sophisticated analysis methods, such as those capble of directly inferring transfer functions from continuum reverberation mapping data (e.g., \citealt{Horne1994, Li2016MICA, Li2021, Anderson2021}) or performing frequency-resolved lag analysis (e.g., \citealt{Zoghbi2021, Cackett2022}), will aid in distinguishing these models. 

\section*{Acknowledgements}

Y.R.L. acknowledges financial support from the the National Natural Science Foundation of China (12273041), from the National Key R\&D Program of China (2023YFA1607904), from the Youth Innovation Promotion Association CAS, and from the China Manned Space Project (CMS-CSST-2025-A07). J.M.W. acknowledges financial support from the National Key R\&D Program of China (2021YFA1600404) and from the National Natural Science Foundation of China (12333003).
R.V. acknowledges financial support from the STFC studentship ST/Y509589/1.
\section*{Data Availability}
The raw LCO and {\it Swift} data of PG 1302-102 underlying this work can be downloaded from the LCO archive at \url{http://archive.lco.global} and the {\it Swift} archive at \url{https://www.swift.ac.uk/}, respectively. The reduced LCO and {\it Swift} photometric light-curve data of PG 1302-102 are available in \cite{Liu2024}. The other public photometric data of PG 1302-102 can be downloaded from the CRTS archive at~(\url{http://nesssi.cacr.caltech.edu/DataRelease/}), ZTF archive at~(\url{https://www.ztf.caltech.edu/ztf-public-releases.html}) and ASAS-SN archive at~(\url{https://www.astronomy.ohio-state.edu/asassn/index.shtml}). 
The SED data of PG 1302-102 can be retrieved from the VizieR catalogues (\url{http://vizier.u-strasbg.fr/vizier/sed/}). The processed data underlying this work are available on reasonable request from the authors.




\bibliographystyle{mnras}
\bibliography{refs.bib} 




\appendix

\section{The Derivation for the responsivity-weighted transfer function}\label{app:A}
In this appendix, we present a derivation for the responsivity-weighted transfer function. According to Equation~(\ref{flux}), the observed flux at a wavelength can be written as  
\begin{equation}\label{eqn_flux}\begin{split}
    F(\lambda,t)
    &= F_{\rm \mu}(\lambda) +F_{\rm \sigma}(\lambda)\int\psi(\lambda, \tau)X(t-\tau)d\tau\\
    &\propto\iint B\left[\lambda,T(t-\tau_{\rm s})\right]r\mathrm{d}r\mathrm{d}\theta.
\end{split}
\end{equation}
Differentiating $F(\lambda, t)$ with respect to $X(t)$ gives
\begin{equation}
    \begin{split}
        \frac{\partial F(\lambda,t)}{\partial X(t)} &=F_{\rm \sigma}(\lambda) \psi(\lambda, t) \\
        &\propto\iint \frac{\partial B } { \partial T }\frac{\partial T}{\partial L_{\rm b}} \frac { \partial L_{\rm b}(t-\tau_{\rm s}) } { \partial X(t-\tau_{\rm s}) } \delta(t - \tau_{\rm s}) r\mathrm{d}r\mathrm{d}\theta.
\end{split}
\end{equation}
After accounting for the normalization, we can directly obtain the expression for responsivity-weighted transfer function in Equation~(\ref{resp}).
By defining
\begin{equation}
x=hc/\lambda kT,
\end{equation}
the transfer function $\psi(\lambda, \tau)$ can be simplified into 
\begin{equation}\begin{split}
    \psi(\lambda, \tau) &\propto
     \iint \frac{B}{L_{\rm b}}\frac{T_{\rm b}^4}{T^4}\frac{xe^x}{e^x-1}\delta(\tau - \tau_{\rm s}) r\mathrm{d}r\mathrm{d}\theta ,\\
\end{split}\end{equation}
where $T_{\rm b}$ is defined by Equation~(\ref{Tb}).

For a face-on ($i=0$) disk surrounding a single SMBH, there exits an analytical expression 
\begin{equation}
\psi(\lambda, \tau) \propto \frac{B}{L_{\rm b}}\frac{T_{\rm b}^4}{T^4}\frac{xe^x}{e^x-1}\sqrt{r^2+h_{\rm s}^2},
\end{equation}
where 
\begin{equation}
r=\sqrt{c^2\tau^2-2c\tau h_{\rm s}}.
\end{equation}

\section{Period Search in the Long-term Variations of PG1302-102}\label{app:period}
\begin{figure*}
\centering
    \includegraphics[width=0.9\textwidth]{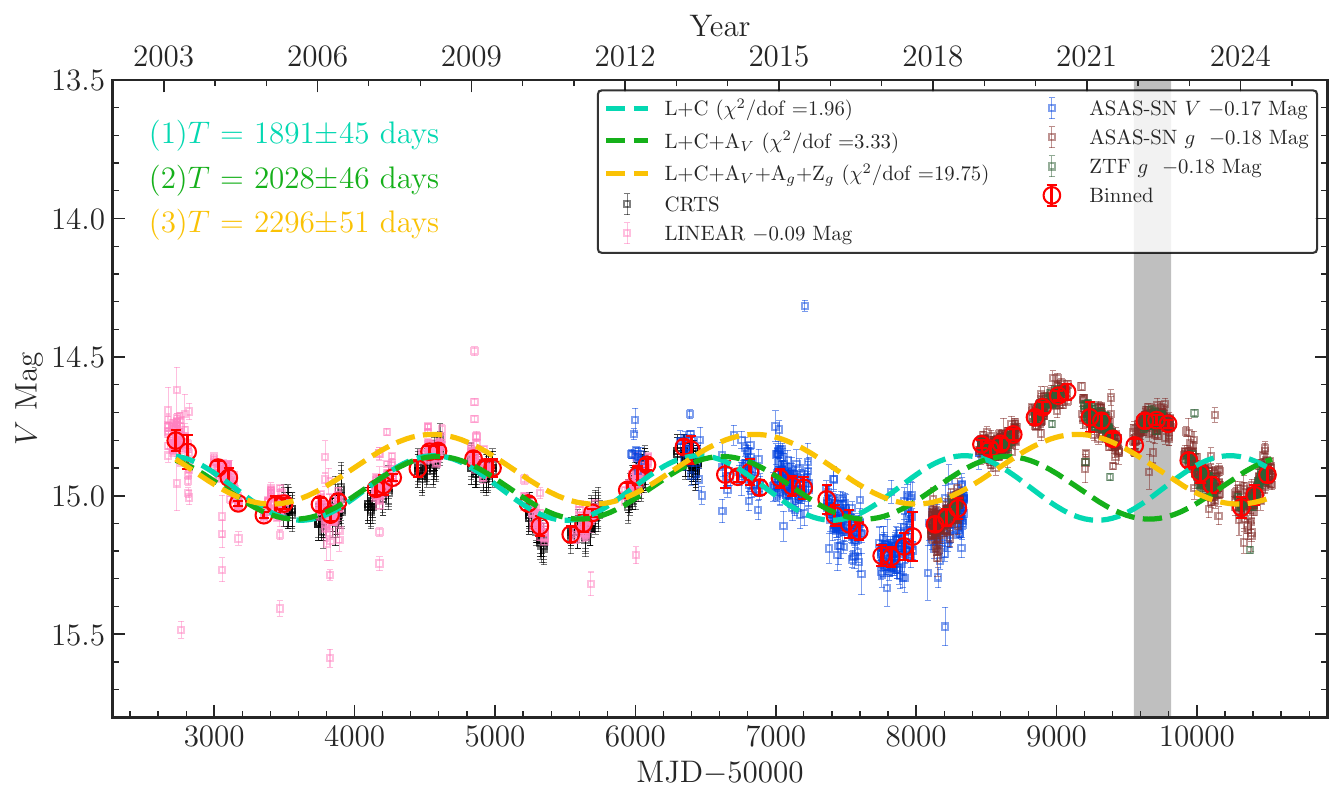}
    \caption{The long-term optical light curve of PG1302-102, compiled by merging the following photometric archival datasets L - LINEAR (pink), C - CRTS (black), A$_{V}$ - ASAS-SN $V$ (blue), A$_{g}$ - ASAS-SN $g$ (dark red), and Z$_{g}$ - ZTF $g$(dark green). The red open circles represent the binned light curve. The dashed lines represent the sinusoidal fits to the binned light curve, with a cyan color for L+C datasets, green for L+C+A$_{V}$ datasets, and orange for L+C+A$_{V}$+A$_{g}$+Z$_{g}$ datasets. The corresponding best-fit periods (in observed frame) are outlined in the top left corner. The gray vertical shaded band denotes the period of the continuum reverberation mapping dataset (MJD 59551-59817).}
    \label{fig:period}
\end{figure*}

\cite{Graham2015} firstly reported the periodic variation signals of PG 1302-102, using the photometric data from Catalina Real-time Transient Survey (CRTS; \citealt{Drake2009}) and Lincoln Near-Earth Asteroid Research (LINEAR; \citealt{Sesar2011}) from MJD 52700 to 56400.
PG 1302-102 was also photometrically monitored by the ASAS-SN survey since MJD 55972 and the ZTF survey since MJD 58203. We compile these photometric data and combine them with the CRTS and LINEAR data. Following \cite{Liu2018}, we offset the ASAS-SN V and LINEAR photometry by -0.17 mag and -0.09 mag, respectively, to align with the CRTS photometry. For the ASAS-SN $g$-band and ZTF $g$-band data, we first use the package \texttt{PyCALI} (\citealt{Li2014}) to intercalibrate them by adopting ZTF $g$ as the reference. The $g$-band photometry is then transformed to $V$-band photometry using  $V = g - 0.03 - 0.52(g-r)$~\citep{Jester2005}.  
The final light curve is plotted in Figure~\ref{fig:period}. To alleviate the influences of the irregular sampling on sinusoidal fits, the light curve is further rebinned every 100 days. The magnitude of each binned point is assigned by the averaged magnitudes with inverse-variance weights and the uncertainty is assigned by the standard deviation of the data points in each bin.

We perform sinusoidal fitting to the binned light curve using the following three cases of datasets, 1) LINEAR+CRTS; 2) LINEAR+CRTS+ASAS-SN $V$, and 3) the whole datasets. The first two cases yield an observed period of 1891$\pm$45 and 2028$\pm$46 days, respectively, well consistent with the results of \cite{Graham2015} and \cite{Liu2018}. The third case yields an observed period $P_{\rm B}=2296 \pm 51$ days, slightly longer than these of the first two cases. This is as expected since a simple visual inspection of Figure~\ref{fig:period} can clearly find that both the sinusoidal period and amplitude appear to be gradually increasing with time.

\section{Fitting the ICCF time lags of PG 1302-102 with the SMBHB model}\label{app:ICCF}
\begin{figure*}
\centering
    \includegraphics[width=0.7\textwidth]{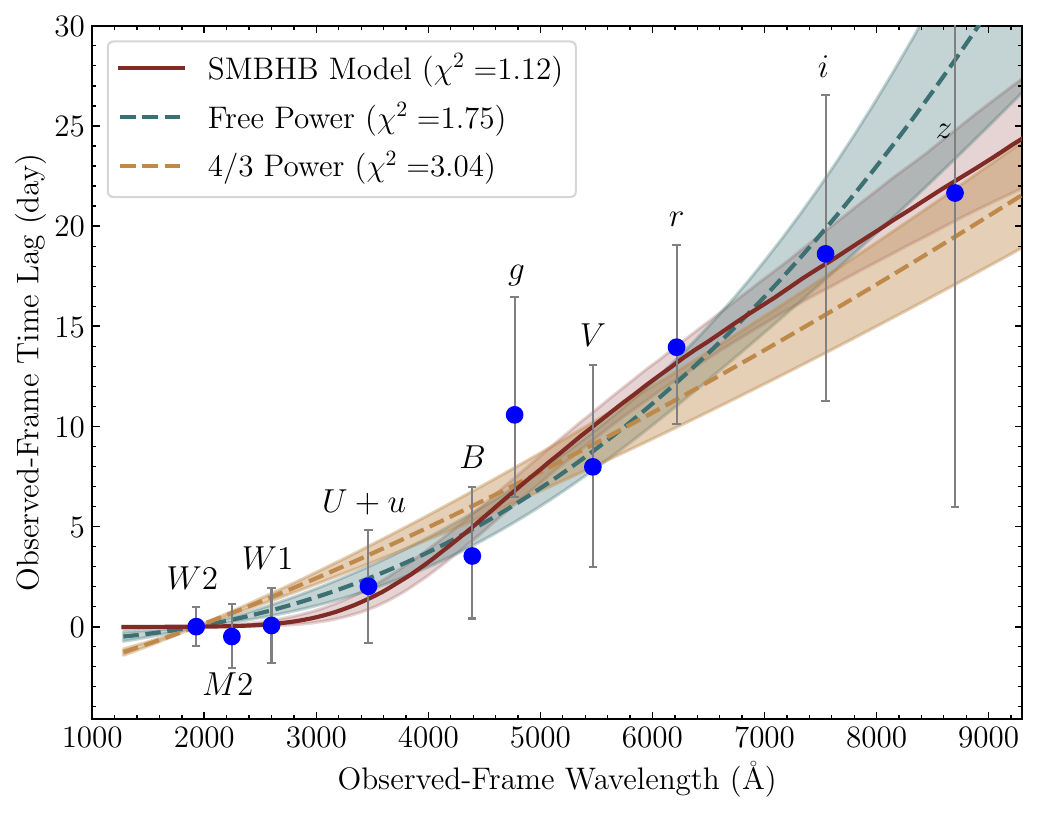}
    \caption{Same as Figure~\ref{fig:1302fit}, but for using the ICCF time lags.}
    \label{fig:1302fit_ICCF}
\end{figure*}

\begin{figure*}
\centering
    \includegraphics[width=0.95\textwidth]{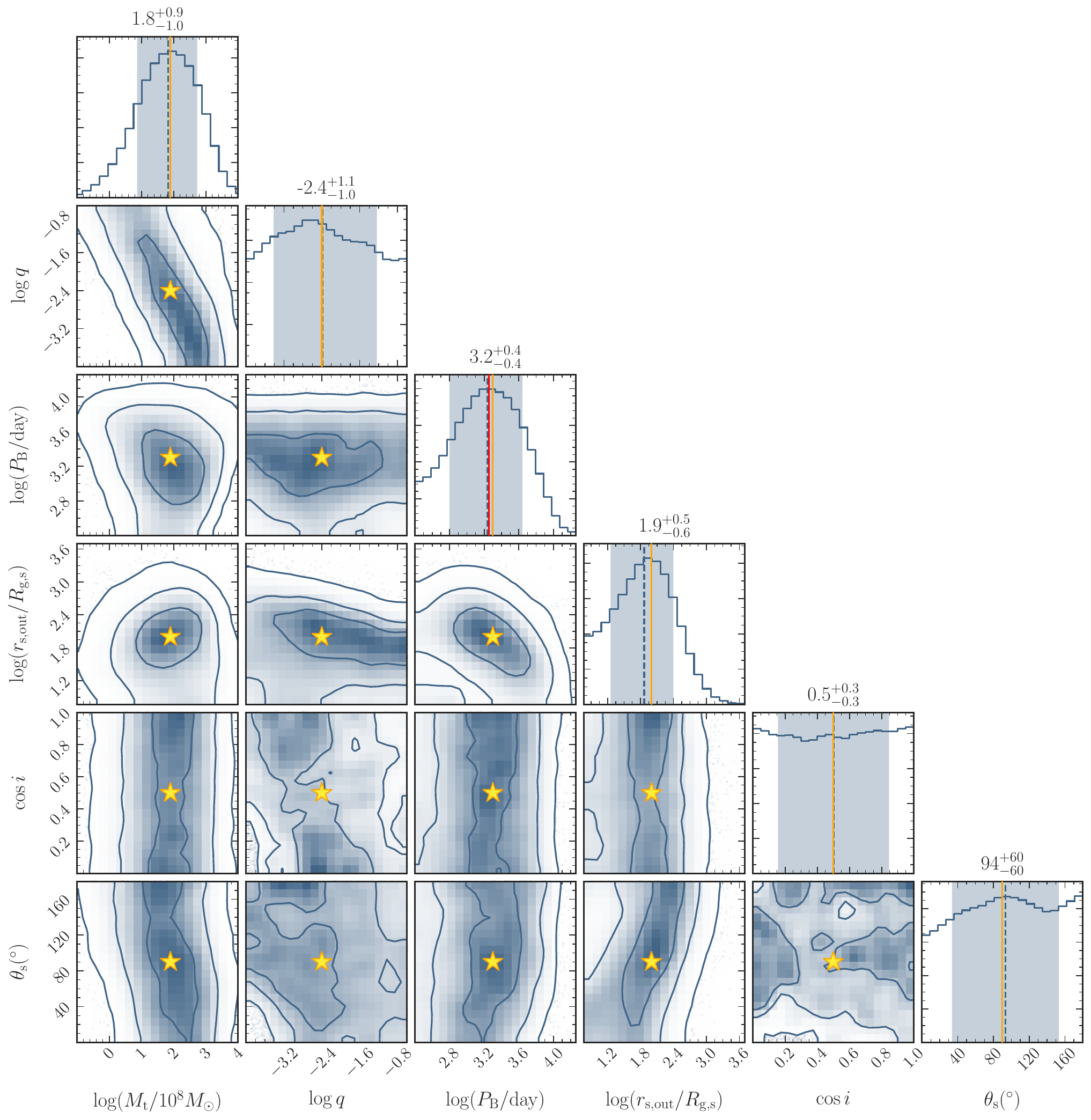}
    \caption{The posterior distribution of the orbital parameters obtained by fitting the ICCF time lags. In the diagonal panels, the blue vertical dashed lines represent the median value and the shaded band represents the 68.3\% confidence interval. The red vertical solid line marks $\log (P_{\rm B}/{\rm day})=3.25$. The orange vertical lines and stars mark the medians of the posterior distribution obtained using the \texttt{MICA} time lags in Figure~\ref{fig:1302fit}. The contours are at $1\sigma$, $2\sigma$, and $3\sigma$ levels.
    }
    \label{fig:corner_ICCF}
\end{figure*}

In Figures~\ref{fig:1302fit_ICCF} and \ref{fig:corner_ICCF}, we show the SMBHB model fit to the ICCF time lags of PG 1302-102 and the obtained posterior distributions of the model parameters, respectively. For the sake of comparison, in Figure~\ref{fig:1302fit_ICCF}, we superimpose the power-law fits in two cases, one with the power-law index $\beta$ free and the other with $\beta$ fixed to $4/3$ (see Equation~\ref{eqn_powerlaw}). Because of the large uncertainties in the ICCF time lags, the SMBHB model and two cases of the power-law model give similar goodness of fit in light of $\chi^2$. The inferred SMBHB model parameters are listed in Table~\ref{tab:best-fit-para}. For the power-law model with $\beta=4/3$, the inferred $\tau_0=2.36 \pm 0.29$ days. For the power-law model with free $\beta$, the inferred $\tau_0=0.61 \pm 0.37$ days and $\beta=2.40 \pm 0.48$. 

\section{The Spectral Energy Distribution of PG 1302-102}\label{app:sed}
\begin{figure*}
    \centering 
    \includegraphics[width=0.9\linewidth]{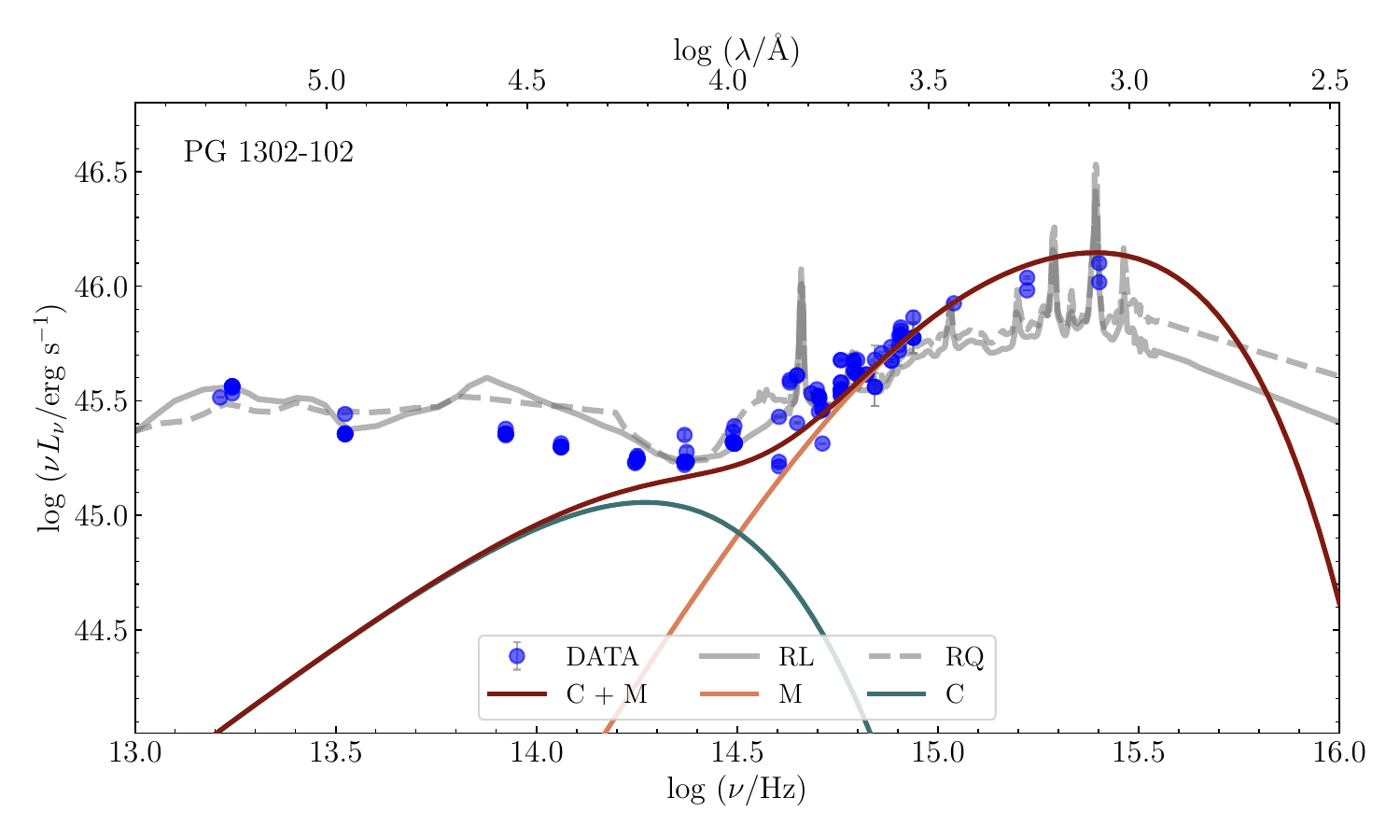}
    \caption{The spectral energy distribution of PG 1302-102. The data points are retrieved from the VizieR catalogues (\url{http://vizier.u-strasbg.fr/vizier/sed/}). 
    The green, orange, and red lines show the predicted SEDs of the circumbinary disk (C), the mini-disk (M), and their combination (C+M), respectively, which are are calcuated using the SMBHB parameters $\log(M_{\rm t}/10^8M_\odot)=2.26$, $\log q=-0.79$, $\log(P_{\rm B}/{\rm yr})=3.43$, $\log(r_{\rm s, out}/R_{\rm g, s})=1.52$, $\cos i=0.24$, and $\theta_{\rm s}=125^\circ$. These parameters are selected from the posterior samples in Figure~\ref{fig:1302fit} by minimizing the $\chi^2$. 
    The grey solid and dashed lines represent the composite SEDs of radio-quiet (RQ) and radio-loud (RL) quasars (\citealt{Shang2011}), repsectively, which are aligned to the same luminosity of PG~1302-102 at SDSS $i$ band ($\lambda=7625$~{\AA}). }
    \label{fig:sed}
\end{figure*}

We retrieve archival photometric data of PG 1302-102 using the Vizier tool\footnote{\url{http://vizier.u-strasbg.fr/vizier/sed/}} with a search radius of 3{\arcsec}.  
The photometry is corrected for the Galactic extinction using the dust map of \cite{Schlegel1998} with the reddening law of \cite{Schlafly2011}. Figure~\ref{fig:sed} shows these SED data of PG 1302-102. For comparison, we superimpose the theorectical SEDs calculated from the SMBHB model using the parameters selected from the posterior samples in Figure~\ref{fig:1302fit} that miminize the $\chi^2$. Here, $\chi^2$ is calculated over the data with $\log(\lambda/\angstrom)<4.2$, excluding the infrared data that might be dominated by emissions from the outer torus. 

The disks' luminosities are calculated as
\begin{equation}
L(\lambda)=4\pi \cos i \iint B(\lambda, T) r dr d\theta,
\end{equation}
where the surface temperature $T$ is given by Equations~(\ref{T}) and (\ref{Tb}) with the corona luminosity $L_{\rm b}$ fixed to the long-term mean luminosity $L_\mu$. The parameter values are $\log(M_{\rm t}/10^8M_\odot)=2.26$, $\log q=-0.79$, $\log(P_{\rm B}/{\rm yr})=3.43$, $\log(r_{\rm s, out}/R_{\rm g, s})=1.52$, $\cos i=0.24$, and $\theta_{\rm s}=125^\circ$.

\bsp	
\label{lastpage}
\end{document}